\documentclass[12pt]{article}

\usepackage{epsfig}
\usepackage{latexsym}
\usepackage{color}
\usepackage{cite}
\usepackage{amsmath}
\usepackage{amssymb}
\newcommand{\fun}{$F_2^{\gamma}(x,Q^2)$ }

\newcommand{\gam}{^{\gamma}}
\newcommand{\gamn}{^{\gamma,n}}
\newcommand{\fund}{$F_2^{\gamma}(x,Q^2)$}
\newcommand{\be}{\begin{equation}}
\newcommand{\ee}{\end{equation}}
\newcommand{\ba}{\begin{eqnarray}}
\newcommand{\ea}{\end{eqnarray}}
\newcommand{\etal}{{\it et al.}}
\newcommand{\dis}{DIS$_{\gamma}$ }
\newcommand{\diss}{DIS$_{\gamma}$}
\newcommand{\dism}{\mathrm{DIS}_{\gamma}}
\newcommand{\ms}{${\mathrm{\overline{MS}}}$ }
\newcommand{\mss}{${\mathrm{\overline{MS}}}$}
\newcommand{\msm}{{\mathrm{\overline{MS}}}}
\newcommand{\alsq}{\alpha_s(Q^2)}
\newcommand{\alsm}{\alpha_s(Q_0^2)}
\newcommand{\pl}{\mathrm{pl}}
\newcommand{\had}{\mathrm{had}}
\newcommand{\bz}{\beta_0}

\def\ie{{i.e.} }

\begin{document}

\bibliographystyle{unsrt}

\begin{center}
\vspace{1cm}
\Large{A New 5 Flavour NLO Analysis and Parametrizations of Parton
Distributions of the Real Photon}\\
\vspace{1cm}
\large{F.Cornet$^a$, P.Jankowski$^b$ and M.Krawczyk$^{b}$}\\
\vspace{0.4cm}
\normalsize{
{$^a$\sl Departamento de F\'{\i}sica Te\'orica y del Cosmos, Universidad de 
Granada,\\
Campus de Fuente Nueva, E-18071, Granada, Spain}\\
\vspace{0.2cm}
{$^b$\sl Institute of Theoretical Physics, Warsaw University,\\
ul. Ho\.za 69, 00-681 Warsaw, Poland}\\
}
\end{center}
\vspace{2cm}

\begin{abstract}
New, radiatively generated, NLO quark $(u,d,s,c,b)$ and gluon densities in a
real, unpolarized photon are presented. We perform three global fits, based on 
the NLO DGLAP evolution equations for $Q^2>1$ GeV$^2$, to all the available 
structure function \fun data. As in our previous LO analysis we utilize two 
theoretical approaches. Two models, denoted as FFNS$_{CJK}$1 \& 2 NLO, adopt 
the so-called Fixed Flavour-Number Scheme for calculation of the heavy-quark 
contributions to \fund, the CJK NLO model applies the ACOT($\chi$) scheme. We 
examine the results of our fits by a comparison with the LEP data for the $Q^2$
dependence of the $F_2\gam$, averaged over various $x$-regions, and the 
$F_{2,c}\gam$. Grid parametrizations of the parton densities for all fits are 
provided.
\end{abstract}
\newpage


\section{Introduction}

In this work we extend the LO QCD analysis of the structure function \fun of 
the unpolarized real photon \cite{cjkl,cjk,jank}, to the Next-to-Leading order
(NLO). As before we are especially interested in the description of the charm-
and beauty-quark contributions to \fund. We adopt two approaches, the 
FFNS$_{CJK}$ and CJK ones, as discussed in \cite{cjkl} and (with some 
modifications) in \cite{cjk}. In the first model, referred to as 
FFNS$_{CJK}$1 NLO, we adopt a standard approach in which a heavy quark, $h$,
contributes to the photon structure function only through the 'direct' 
(Bethe-Heitler) $\gamma^* +\gamma \to h+\bar h$ process and the process with 
'resolved photon' $\gamma^* +G\gam \to h+\bar h$. In the FFNS$_{CJK}$2 NLO
model we include in addition $\gamma^* +G\gam \to h+\bar h+G$ and 
$\gamma^* +q\gam(\bar q\gam)\to q(\bar q)+h+\bar h$ processes\footnote{These 
two FFNS$_{CJK}$1 \& 2 models differ from the corresponding models introduced 
in our LO analysis.}. In the third model, CJK NLO, we use the ACOT($\chi$) 
scheme \cite{acot} in which, apart from the direct and resolved photon 
$O(\alpha_s)$ contributions\footnote{Inclusion of the $O(\alpha_s^2)$ terms 
in the CJK NLO approach is beyond the scope of this work.}, we deal with the 
heavy-quark densities $q_h(x,Q^2)$. The subtraction of the double counted 
terms as well as introduction of a new kinematic variable $\chi_h$ improve 
description of \fun in both regions: below and above the heavy-quark 
thresholds.

In \cite{cjkl,cjk,jank} we followed the idea of the radiatively generated 
parton distributions introduced by the GRV group first to describe nucleon 
\cite{grvn} and pion \cite{grvp} and later used for the real \cite{grv92}, 
\cite{grs} and virtual photon \cite{grs}, \cite{grst}. In our LO analyses
\cite{cjkl,cjk,jank} we used $Q_0^2=0.25$ GeV$^2$ and adopted the input 
distributions of the valence-type form introduced in \cite{grv92} neglecting 
sea-quark densities. In the case of the NLO analysis we found that the starting
scale of the evolution appears to tend to higher values, 
$Q_0^2\approx 0.7$ GeV$^2$ (similarly as in \cite{klasen}). We tested the 
assumption on vanishing sea-quark densities at $Q_0^2$ by performing additional
fits.

All our global fits are performed at $Q^2>1$ GeV$^2$ utilizing the set of all 
available \fun data.

We test our results for \fun by comparing them with the LEP data which were not
used in the fits: for the $Q^2$ dependence of the $F_2\gam$, averaged over 
various $x$-regions, and for the $F_{2,c}\gam$ data.

Our paper is divided into six parts. In section 2 we shortly recall basic
schemes applied to calculation of the heavy-quark production in the Deep 
Inelastic Scattering on the real-photon target. Section 3 is devoted to the 
description of the FFNS$_{CJK}$ NLO models of the \fund. Next, the CJK NLO 
model based on the ACOT($\chi$) scheme \cite{acot} is presented in detail. In 
section 4 we describe the solutions of the DGLAP evolution equations in all
models and assumptions for the input parton densities. Results of the 
global fits are discussed and compared with the \fun, $F_2^{\gamma}(Q^2)$ and 
$F_{2,c}\gam$ data in the fifth section of the paper. We end our paper by 
presenting in the Appendix the technical details of the calculation. The 
parton distributions resulting from our fits have been parametrized on 
the grid.


\section{Production of heavy quarks in DIS$_{e\gamma}$}

\begin{figure}[t]
\begin{center}
\epsfxsize=20pc
\epsfbox{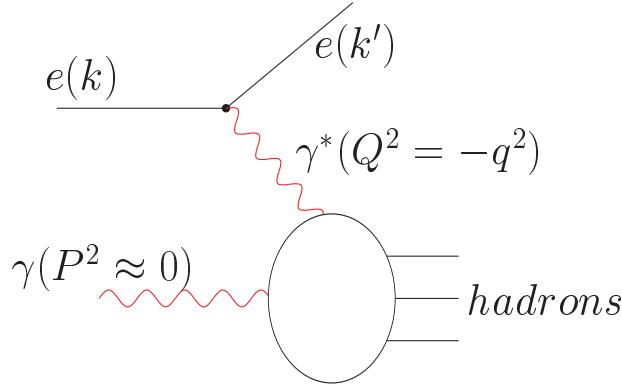}
\end{center}
\caption{\small Deep Inelastic Scattering on quasi real photon
($P^2\approx 0$), $e+\gamma \to e + hadrons$.}
\label{egeh}
\end{figure}

The Deep Inelastic Scattering on a photon (DIS$_{e\gamma}$) allows to 
measure the structure function $F_2^{\gamma}$, and others, via the process
\be
e + \gamma \to e + hadrons,
\ee
presented in the figure \ref{egeh}. The differential cross-section for this 
process is given by:
\be
\frac{d\sigma^{e\gamma\to eX}}{dxdy} = \frac{2\pi\alpha^2}{xyQ^2}
[(1+(1-y)^2)F_2^{\gamma}(x,Q^2)-y^2F_L^{\gamma}(x,Q^2)],
\ee
where $x$ and $y$ are the standard DIS$_{e\gamma}$ variables, defined through 
the virtuality $Q^2$ of the probing photon and momenta of interacting 
particles shown in figure \ref{egeh},
\be
x = \frac{Q^2}{2p\cdot q}, \quad \quad y = \frac{p\cdot q}{p\cdot k},
\ee
and $F_L^{\gamma}(x,Q^2)$ is the so-called longitudinal structure function 
corresponding to the exchange of the longitudinally polarized $\gamma^*$. For 
the presently kinematically accessible values of $y$ the $F_L^{\gamma}(x,Q^2)$
contribution is negligible.

Each of partons (quarks and gluon) of the real photon contributes to the \fun 
proportionally to the corresponding partonic cross-section. The calculation of 
the light $u$-, $d$-, $s$-quark and gluon cross-sections is straightforward.

Various schemes used in the calculation of the heavy-quark production have 
been described in \cite{cjkl}. They are also explained in the vast literature, 
\cite{acot}, \cite{csnc} and \cite{csnb}. Let us shortly recall the main facts
on the FFNS and ACOT($\chi$) approaches applied in this analysis.

There exist two standard schemes in a calculation of a heavy-quark, $h$, 
production at the hard scale $\mu$. In both of them the light quarks $u,d$ and
$s$ are treated as massless because for them $\mu \gg \Lambda_{QCD} > m_q$, 
where $m_q$ is a light-quark mass. In the case of the DIS process the hard 
scale $\mu$ is identified with the virtuality of the probing photon, 
$\mu^2=Q^2$. In the massive Fixed Flavour-Number Scheme (FFNS) heavy $c$- and 
$b$-quarks are treated differently: the massive charm and beauty quarks 
produced in hard subprocesses $\gamma^* +\gamma \to h+\bar h$ can only appear 
in the final state. In the second, massless scheme, called the Zero-mass 
Variable Flavour-Number Scheme (ZVFNS) heavy-quark distributions appear alike 
the light-parton densities. When $Q^2$ is larger than a threshold associated 
with a heavy quark (usually taken as $Q^2=m_h^2$), this quark is considered as
an extra massless parton in addition to the three light quarks. This way, the 
number of different types of quarks (flavours) increases with increasing 
$Q^2$. One introduces a notion of ``active quarks'', for which the condition 
$Q^2>m_q^2$ is fulfilled and which can be treated as massless partons of the 
probed real photon. When active, their densities $q_i(x,Q^2)$ differ from zero,
otherwise $q_i(x,Q^2)=0$.

The above standard schemes are considered to be reliable in different $Q^2$ 
regions. The FFNS looses its descriptive power when $Q^2 \gg m_h^2$, on the 
other hand the ZVFNS does not seem appropriate if $Q^2 \approx m_h^2$. To 
obtain a prescription working for all hard scales one needs to combine 
features of both schemes. There exist the whole set of the approches, generally
denoted as General-mass Variable Flavour-Number Schemes (GVFNS), which in the 
$Q^2\to m_h^2$ and $Q^2\to \infty$ limits reproduce the behaviour of the FFNS 
and ZVFNS schemes respectively. Their recent realization is the so-called 
ACOT($\chi$) scheme introduced for the proton structure function in 
\cite{acot} and used by us in our LO analysis \cite{cjkl,cjk,jank}.

In the ACOT($\chi$) scheme, all contributions to the considered process, which
would be included separately in the ZVFNS and FFNS, are taken into account. Such
procedure requires proper subtraction of double counted contributions, i.e. the 
large logarithms which appear in the massive heavy-quark Wilson coefficients 
(the FFNS scheme) but are also resummed in the $q_h(x,\mu^2)$ densities. That 
way one obtains a scheme in principle appropriate in the whole $Q^2$ range. 
Still, another problem emerges. In any process the kinematical threshold for the
heavy-quark production is given by the total energy of that process, $W$. 
Obviously $W$ must be greater than two masses of the heavy quark in hand, 
$W>2m_h$, so in the DIS$_{e\gamma}$ case where the hard scale is $Q^2$, we have 
$W^2=(1-x)Q^2/x>4m_h^2$. As $\mu^2=Q^2$ the ZVFNS condition $\mu>m_h$ (or any 
similar) for treating a heavy quark $h$ as a massless parton is not correct. It 
may happen (for small enough $x$) that $q_h(x,Q^2)=0$ in the kinematically 
allowed region $W>2m_h$. On the other hand, we can also obtain nonzero 
heavy-quark densities in the kinematically forbidden ($x,Q^2$) region. The 
ACOT($\chi$) scheme solves this problems through introduction of a new variable 
$\chi_h \equiv x(1+4m_h^2/Q^2)$ which replaces in the ZVFNS the Bjorken $x$ as 
an argument of the heavy-parton density.

In the next sections we will present the realizations of the FFNS and 
ACOT($\chi$) schemes for the calculation of the photon structure 
$F_2^{\gamma}$ in the Next-to-Leading order of QCD, namely our FFNS$_{CJK}$ NLO
and CJK NLO models.


\section{Description of the $F_2^{\gamma}$ in the FFNS$_{CJK}$ NLO}

This section is devoted to the FFNS$_{CJK}$1 \& 2 NLO models, which are the 
realizations of the FFNS approach for the photon structure $F_2^{\gamma}$ in 
the Next-to-Leading Order of QCD. The only difference between this two models
is following: the FFNS$_{CJK}$NLO 2 approach includes additional higher order 
contributions to $F_2^{\gamma}$ neglected in the FFNS$_{CJK}$NLO 1 model.

First, we describe the light-quark contributions to \fun which are common for
the both analysed models.  Next, we discuss the heavy-quark contributions for 
the FFNS$_{CJK}$NLO approach. Further, we recall the \dis factorization scheme
used in the model and resume by giving the final formulea for the \fun in the 
FFNS$_{CJK}$ NLO models.


\subsection{Light-quark contributions \label{lightsect}}

In the Next-to-Leading Order logarithmic approximation of QCD or, in short, in 
NLO QCD the light-quark contributions to the photon structure function \fun 
can be written in terms of light-quark ($u,d$ and $s$) densities 
$q_i\gam (x,Q^2)$ as follows:
\ba
&& \frac{1}{x}F_2\gam(x,Q^2)|_{light} =
\sum_{i=1}^3 e_i^2 \left\{ (q_i\gam+\bar q_i\gam)(x,Q^2) + 
e_i^2 \frac{\alpha}{2\pi}C_{2,\gamma}^{(0)}(x) \right.  \label{lightf} \\
&+& \left. \frac{\alpha_s(Q^2)}{2\pi}\int_x^1\frac{dy}{y}\left[ 
(q_i\gam+\bar q_i\gam)(y,Q^2) C^{(1)}_{2,q}(\frac{x}{y}) 
+ G\gam(y,Q^2)C^{(1)}_{2,G}(\frac{x}{y}) \right] \right\}. \nonumber
\ea
Note, that $q_i\gam(x,Q^2) = \bar q_i\gam(x,Q^2)$. The $C_{2,j}^{(i)}(x)$ 
functions are the $O(\alpha_s^i)$ order terms of the hadronic (Wilson) 
coefficient functions ($j=\gamma, q, G$)
\be
C_{2,j}(x,Q^2) = C_{2,j}^{(0)}(x) + \frac{\alpha_s(Q^2)}{2\pi}C_{2,j}^{(1)}(x)
+ \cdots .
\ee
Each of them is related to the hard process of a parton originating 
from the real photon (for instance quark for $C_{2,q}(x)$) interacting
with the virtual photon. The coefficients appearing in Eq. (\ref{lightf}) are 
listed in table \ref{WCL}. The formulae for the $C_{2,j}^{(i)}(x)$ functions 
up to the $O(\alpha_s^2)$ order can be found in \cite{LRSN}, see also the 
Appendix.

\begin{table}[h]
\begin{center}
\begin{tabular}{|c|@{} p{0.1cm} @{}|c|c|}
\hline
  order         &&       hard (parton) process        & coefficient function \\
\hline
\hline
$O(\alpha_s^0)$ && $\gamma^*+q\gam(\bar q\gam) \to q(\bar q)$ & $C_{2,q}^{(0)}(x)=\delta(1-x)$ \\
\cline{3-4}
                && $\gamma^*+\gamma \to q + \bar q$   & $C_{2,\gamma}^{(0)}(x)$ \\
\hline
\hline
$O(\alpha_s^1)$ && $\gamma^*+q\gam(\bar q\gam) \to q(\bar q) + G$ & $C_{2,q}^{(1)}(x)$ \\
\cline{3-4}
                && $\gamma^*+G\gam \to q + \bar q$        & $C_{2,G}^{(1)}(x)$ \\
\hline
\end{tabular}
\caption{\small Hard parton processes and the corresponding Wilson coefficient
functions for light quarks.}
\label{WCL}
\end{center}
\end{table}

The evolution of the parton densities in $\ln Q^2$ is governed by the 
inhomogeneous DGLAP equations \cite{ap}. In the NLO we have:
\ba
&& \frac{dq_i\gam(x,Q^2)}{d\ln Q^2} = \frac{\alpha}{2\pi} e_i^2k_{q}(x,Q^2) \\
&+& \frac{\alpha_s(Q^2)}{2\pi} \int_x^1\frac{dy}{y} \left\{ 
\sum_{k=1}^{N_f} \left[ q_k\gam(y,Q^2)P_{qq}(\frac{x}{y},Q^2)
                  + \bar q_k\gam(y,Q^2)P_{q\bar q}(\frac{x}{y},Q^2) \right]
+ G\gam(y,Q^2)P_{qG}(\frac{x}{y},Q^2) \right\}, \nonumber \label{DGLAP1} \\
&& \frac{dG\gam(x,Q^2)}{d\ln Q^2} = \frac{\alpha}{2\pi} k_G(x,Q^2) \\
&+& \frac{\alpha_s(Q^2)}{2\pi} \int_x^1\frac{dy}{y} \left\{ 
\sum_{k=1}^{N_f} \left[ q_k\gam(y,Q^2)P_{Gq}(\frac{x}{y},Q^2) 
                  + \bar q_k\gam(y,Q^2)P_{G\bar q}(\frac{x}{y},Q^2) \right]
+ G\gam(y,Q^2)P_{GG}(\frac{x}{y},Q^2)\right\} \nonumber \label{DGLAP2}
\ea
where $N_f$ is a number of active quarks. Here $N_f=3$ as all heavy-quark 
densities are equal zero in the FFNS models. The functions $P_{ij}(x,Q^2)$
and $k_i(x,Q^2)$ are the NLO splitting functions
\ba
k_i(x,Q^2) &=& k_i^{(0)}(x)+\frac{\alpha_s(Q^2)}{2\pi}k_i^{(1)}(x), \\
P_{ij}(x,Q^2) &=& P_{ij}^{(0)}(x)+\frac{\alpha_s(Q^2)}{2\pi}P_{ij}^{(1)}(x). 
\nonumber
\ea
The $k_q(x,Q^2)$ and $k_G(x,Q^2)$ are reffered to as the photon-quark and 
photon-gluon splitting functions, respectively. Up to the NLO order 
$P_{Gq}=P_{G\bar q}$. Formulae for $O(\alpha_s)$ and $O(\alpha_s^2)$ splitting
functions can be found for instance in \cite{Stirling}.


\subsection{Heavy-quark contributions \label{heavyffnssect}}

In the FFNS models heavy quarks appear only in the final state of
the process and the density functions $q_h(x,Q^2)$ are equal to zero. The 
charm- and beauty-quark contributions to the structure function are described 
by the heavy-quark coefficient functions $C_{2,j}^{h,(i)}$ related to the hard 
processes as listed in table \ref{WCH}. One obtains
\ba
&& \frac{1}{x}F_2^{\gamma}(x,Q^2)|_{heavy,FFNS} = 
\sum_{h(=c,b)}^2 e_h^2 \left\{ \frac{\alpha_s(Q^2)}{2\pi} 
\int_{\chi_h}^1\frac{dy}{y}G\gam(y,Q^2)C_{2,G}^{h,(1)}(\frac{x}{y},\frac{Q^2}{m_h^2})
+ e_h^2 \frac{\alpha}{2\pi}C_{2,\gamma}^{h,(0)}(x,\frac{Q^2}{m_h^2}) \right\} \nonumber 
\label{heavyf} \\
&& + \sum_{h(=c,b)}^2 \left\{ \left(\frac{\alpha_s(Q^2)}{2\pi}\right)^2 
\int_{\chi_h}^1\frac{dy}{y} \left[
\sum_{i=1}^{3}(q_i\gam+\bar q_i\gam)(y,Q^2)(e_i^2 C_{2,q}^{(2)}(\frac{x}{y},\frac{Q^2}{m_h^2}) + 
e_h^2 C^{h,(2)}_{2,q}(\frac{x}{y},\frac{Q^2}{m_h^2})) \right. \right. \\
&& \left. + e_h^2G\gam(y,Q^2)C_{2,G}^{h,(2)}(\frac{x}{y},\frac{Q^2}{m_h^2}) \right] \left. 
+ \frac{\alpha \alpha_s(Q^2)}{(2\pi)^2} e_h^4 C_{2,\gamma}^{h,(1)}(x,\frac{Q^2}{m_h^2}) 
\right\}, \nonumber
\ea

\begin{table}[h]
\begin{center}
\begin{tabular}{|c|@{} p{0.1cm} @{}|c|c|}
\hline
  order         &&       parton subprocess            & coefficient function \\
\hline
\hline
$O(\alpha_s^0)$ && $\gamma^*+\gamma \to h+\bar h$ & $C_{2,\gamma}^{h,(0)}$ \\
\hline
\hline
$O(\alpha_s^1)$ && $\gamma^*+G\gam \to h+\bar h$      & $C_{2,G}^{h,(1)}$ \\
\cline{3-4}
                && $\gamma^*+\gamma \to h+\bar h+G$ & $C_{2,\gamma}^{h,(1)}$ \\
\hline
\hline
$O(\alpha_s^2)$ && $\gamma^*+G\gam \to h+\bar h+G$    & $C_{2,G}^{h,(2)}$ \\
\cline{3-4}
                && $\gamma^*+q\gam(\bar q\gam) \to q(\bar q)+h+\bar h$ & $C_{2,q}^{h,{(2)}}$, $C_{2,q}^{(2)}$\\
\hline
\end{tabular}
\caption{\small Hard parton processes and the corresponding Wilson coefficient
functions for heavy quarks.}
\label{WCH}
\end{center}
\end{table}

where the parameter $\chi_h = x(1+4m_h^2/Q^2)$ takes care of the proper region 
of the integration.

We consider two FFNS models, FFNS$_{CJK}$1 and FFNS$_{CJK}$2 NLO, where only
in the latter we include the $O(\alpha_s^2)$ and $O(\alpha \alpha_s)$ 
contributions to $F_2\gam$. The reason for considering two models is twofold. 
First, to check the size of the higher order heavy-quark contributions. 
Secondly, to perform a test whether the $O(\alpha_s^2)$ and 
$O(\alpha \alpha_s)$ terms should be included in the calculations as they are 
of higher order than the other NLO contributions.

\begin{figure}[h]
\begin{center}
\epsfxsize=10pc
\epsfbox{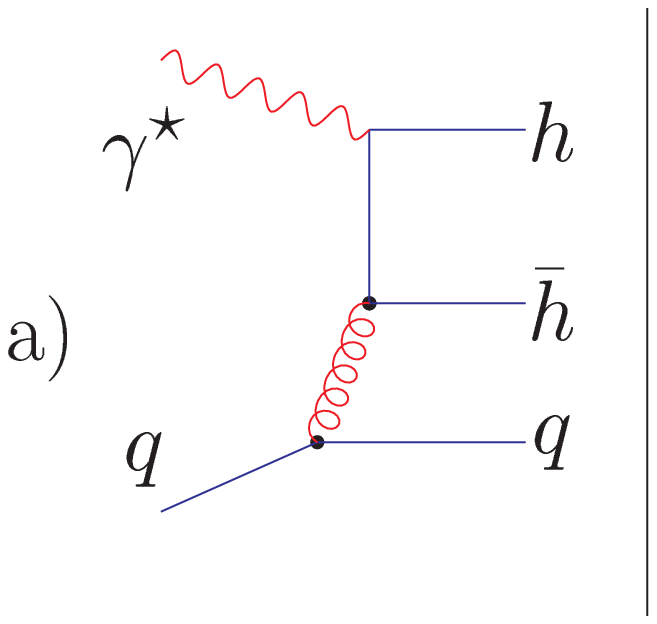}
\epsfxsize=11pc
\epsfbox{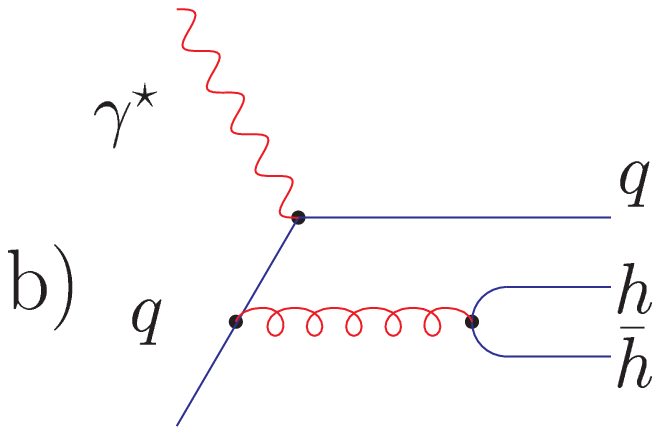}
\caption{\small Diagrams relevant for 
$\gamma^*+q\gam(\bar q\gam) \to q(\bar q)+h+\bar h$ process corresponding to a)
the $C_{2,q}^{h,(2)}$ coefficient and b) the $C_{2,q}^{(2)}$ coefficient.}
\label{c2hqfig}
\end{center}
\end{figure}

Let us notice that among the higer order terms there appear two Wilson 
coefficients originating from the process 
$\gamma^*+q\gam(\bar q\gam) \to q(\bar q)+h+\bar h$, the $C_{2,q}^{h,(2)}$ and 
$C_{2,q}^{(2)}$ (see Fig. \ref{c2hqfig}), the interference term of two diagrams
(proportional to $e_h e_q$) gives 0 according to \cite{LRSN}.

\subsection{The \ms and \dis factorization schemes \label{dis1sect}}

All the splitting and coefficient functions introduced in former sections
have been derived in the modified minimal subtraction factorization scheme 
(\mss). (They all can be found in the literature, see Appendix A). Let us notice
that the $C_{2,\gamma}^{(0)}(x)$ term in Eq. \ref{lightf}, describing the 
'direct' contribution of the photon to \fun causes a problem in the \ms scheme 
\cite{grvdis}. Firstly, contrary to all other contributions, this term does 
not vanish at the input scale $Q_0^2$. One obtains very different boundary 
conditions for the light-quark contribution in LO and NLO:
\be
\frac{1}{x}F_2\gam(x,Q_0^2)|_{light} = \left\{ \begin{array}{cl}
0 & \mathrm{, LO} \\
\sum_{i=1}^3 e_i^4 \frac{\alpha}{2\pi}C_{2,\gamma}^{(0)}(x) & \mathrm{, NLO}.\\
\end{array} \right.
\ee
This leads to very distinct results over the whole $x$-range (see for instance 
\cite{vogt}). The heavy-quark contributions, $C_{2,\gamma}^{h,(k)}(x,\frac{Q^2}{m_h^2})$, 
for $k=0,1$, of Eq. (\ref{heavyf}) do not cause a similar problem as the 
$O(\alpha_s^0)$ term appears already in the LO calculation and the 
$C_{2,\gamma}^{h,(1)}(x,\frac{Q^2}{m_h^2})$ coefficient is only its $O(\alpha_s^1)$ 
correction.

Secondly, the $C_{2,\gamma}^{(0)}(x)$ term calculated in the \ms scheme is 
negative and divergent in the large-$x$ region, where
$C_{2,\gamma}^{(0)}(x) \simeq 3[\ln (1-x)-1]$, leading to unphysical
negative values of $F_2\gam$ at $x \to 1$. It is not cured by addition of 
other NLO contributions to the structure function as they all vanish 
at $x=1$ (if a typical choice of the shape of the hadronic input is made).

In a new factorization scheme, denoted as \diss, introduced in \cite{grvdis},
one avoids the troubles connected with the $C_{2,\gamma}^{(0)}(x)$ term by 
absorbing it into the quark distributions $q_i\gam(x,Q^2)$, namely
\ba
q^{\dism}_i(x,Q^2) &=& q^{\msm}_i(x,Q^2) 
+ e_i^2 \frac{\alpha}{4\pi}C_{2,\gamma}^{(0),\msm}(x), \\
C_{2,\gamma}^{(0),\dism}(x) &=& C_{2,\gamma}^{(0),\msm}(x) 
+ \Delta C_{2,\gamma}^{(0)}(x) = 0, \nonumber
\ea
where obviously 
\be
\Delta C_{2,\gamma}^{(0)}(x) = C_{2,\gamma}^{(0),\msm}(x).
\label{delc2q}
\ee
The gluon density and other coefficient functions are unaltered by the above 
redefinition but it leads to a modification of the photon-quark and 
photon-gluon splitting functions appearing in the DGLAP evolution of the 
parton distributions:
\ba
k_q^{(1),\dism}(x,Q^2) &=& k_q^{(1),\msm}(x,Q^2) 
- \frac{1}{2} \int_x^1 \frac{dy}{y} P_{qq}^{(0)}(\frac{x}{y}) 
C_{2,\gamma}^{(0),\dism}(y), \label{modkqg} \\
k_G^{(1),\dism}(x,Q^2) &=& k_G^{(1),\msm}(x,Q^2) 
- \sum_{i=1}^{N_f} e_i^2 
\int_x^1 \frac{dy}{y} P_{Gq}^{(0)}(\frac{x}{y}) C_{2,\gamma}^{(0),\dism}(y). 
\nonumber
\ea

In our analysis we apply the \dis factorization scheme, for simplicity the \dis
superscript will be omitted below, if possible.


\subsection{Formulae for $F_2^{\gamma}$ \label{f2ffnsform}}

\fun in the FFNS$_{CJK}$NLO 1 model reads;
\ba
&& \frac{1}{x}F_2\gam(x,Q^2)|_{FFNS}1 = \label{f2ffns1} \\
&& \sum_{i=1}^3 e_i^2 \left\{ (q_i\gam+\bar q_i\gam)(x,Q^2) 
+\frac{\alpha_s(Q^2)}{2\pi}\int_x^1\frac{dy}{y}\left[ 
(q_i\gam+\bar q_i\gam)(y,Q^2) C^{(1)}_{2,q}(\frac{x}{y}) 
+ G\gam(y,Q^2)C^{(1)}_{2,G}(\frac{x}{y}) \right] \right\} \nonumber \\
&+& \sum_{h(=c,b)}^2 e_h^2 \left\{ \frac{\alpha_s(Q^2)}{2\pi} 
\int_{\chi_h}^1\frac{dy}{y}G\gam(y,Q^2)C_{2,G}^{h,(1)}(\frac{x}{y},\frac{Q^2}{m_h^2})
+ e_h^2 \frac{\alpha}{2\pi}C_{2,\gamma}^{h,(0)}(x,\frac{Q^2}{m_h^2}) \right\}. \nonumber
\ea
In the case of the $F_2\gam(x,Q^2)|_{FFNS}2$ model we include the 
$O(\alpha_s^2)$ and $O(\alpha \alpha_s)$ terms, obtaining
\ba
&& \frac{1}{x}F_2\gam(x,Q^2)|_{FFNS}2 = \frac{1}{x}F_2\gam(x,Q^2)|_{FFNS}1
\label{f2ffns2} \\
&& + \sum_{h(=c,b)}^2 \left\{ \left(\frac{\alpha_s(Q^2)}{2\pi}\right)^2 
\int_{\chi_h}^1\frac{dy}{y} \left[
\sum_{i=1}^{3}(q_i\gam+\bar q_i\gam)(y,Q^2)(e_i^2 C_{2,q}^{(2)}(\frac{x}{y},\frac{Q^2}{m_h^2}) + 
e_h^2 C^{h,(2)}_{2,q}(\frac{x}{y},\frac{Q^2}{m_h^2})) \right. \right. \nonumber \\
&& \left. + e_h^2G\gam(y,Q^2)C_{2,G}^{h,(2)}(\frac{x}{y},\frac{Q^2}{m_h^2}) \right] \left. 
+ \frac{\alpha \alpha_s(Q^2)}{(2\pi)^2} e_h^4 C_{2,\gamma}^{h,(1)}(x,\frac{Q^2}{m_h^2}) 
\right\}. \nonumber
\ea


\section{Description of the $F_2^{\gamma}$ in the CJK NLO model}

This section is devoted to the description of the CJK NLO model based on
the ACOT($\chi$) approach. The approach combines the FFNS scheme introduced in
the former section with the ZVFNS scheme in which heavy quarks are treated in 
the similar way as the light ones.


\subsection{Light-quark contributions}

The light-quark contributions to $F_2^{\gamma}$ are the same in FFNS and ZVFNS
schemes and were already given in Eq. (\ref{lightf}) for the \ms factorization 
scheme and in Eq. (\ref{f2ffns1}) for the \dis scheme. The difference between 
the numerical values of those contributions originates from the difference 
between the number of the active quarks, $N_f$, in the DGLAP equations in both 
approaches. We have $N_f=3$ for the FFNS$_{CJK}$ NLO models, while $N_f=5$ in 
the CJK NLO model.


\subsection{Heavy-quark contributions \label{heavycjksect}}

Apart from the heavy-quark terms which properly describe the $h$-quark 
contributions to \fun, in the corresponding $Q^2 \approx m_h^2$ regions, are 
given by the FFNS formula (\ref{heavyf}). Let us only notice that in the CJK NLO
model case we neglect the $O(\alpha_s^2)$ and $O(\alpha \alpha_s)$ contributions
to $F_2^{\gamma}(x,Q^2)|_{heavy,FFNS}$. They lead to a very complicated 
expression for $F_2^{\gamma}(x,Q^2)$ and their analysis is beyond the scope of
this work. As already discussed, the FFNS scheme looses its descriptive power
if $Q^2 \gg m_h^2$. Therefore, the addition of the ZVFNS terms, the massless 
charm- and beauty-quark distributions, $q_h(x,Q^2)$, in the photon, appropriate
in that region, is needed. They are combined with the light-quark densities 
through the DGLAP equations with $N_f=5$ active quarks. In the ZVFNS approach heavy-quark 
contributions to \fun read
\ba
&& \frac{1}{x}F_2^{\gamma}(x,Q^2)|_{heavy,ZVFNS} = 
\sum_{h(=c,b)}^2 e_h^2 \left\{ (q_h\gam+\bar q_h\gam)(x,Q^2) 
+ e_h^2 \frac{\alpha}{2\pi}C_{2,\gamma}^{(0)}(x)\right. \label{heavyz} \\
&+& \left. \frac{\alpha_s(Q^2)}{2\pi} \int_x^1\frac{dy}{y} 
\left[ (q_h\gam+\bar q_h\gam)(y,Q^2)C^{(1)}_{2,q}(\frac{x}{y})+
G\gam(y,Q^2)C_{2,G}^{(1)}(\frac{x}{y}) \right] \right\}. \nonumber
\ea


\subsection{Subtraction terms}

A simple summing of all the discussed terms, namely
\be
F_2^{\gamma}(x,Q^2) = F_2^{\gamma}(x,Q^2)|_{light} 
+ F_2^{\gamma}(x,Q^2)|_{heavy,FFNS} + F_2^{\gamma}(x,Q^2)|_{heavy,ZVFNS},
\label{f2sum}
\ee
would lead to double counting of some contributions. Let us notice first that 
the ZVFNS subprocess with the direct coupling of the real photon to the 
(massless) heavy quark, term $C_{2,\gamma}^{(0)}(x)$ in Eq. (\ref{heavyz}), 
was also included in the FFNS formula (\ref{heavyf}) as 
$C_{2,\gamma}^{h,(0)}(x,\frac{Q^2}{m_h^2})$. Those two terms originate from 
the same $\gamma^* +\gamma \to h+\bar h$ process, however they are not equal 
as the $C_{2,\gamma}^{(0)}(x)$ term is obtained in a massless approximation. 
The same is true for the $C_{2,G}^{(0)}(x)$ (Eq. (\ref{heavyz})) and 
$C_{2,G}^{h,(0)}(x,\frac{Q^2}{m_h^2})$ (Eq. (\ref{heavyf})) terms arising from 
the $\gamma^*+G\gam \to h+\bar h$ process. Obviously, only one set of the above
terms should be included in the final CJK NLO formula for the \fund. We 
choose to apply the massive $C_{2,\gamma}^{h,(0)}(x,\frac{Q^2}{m_h^2})$ and 
$C_{2,G}^{h,(0)}(x,\frac{Q^2}{m_h^2})$ formulea, as unlike the massless approximations 
those terms properly vanish at the heavy-quark threshold.

The $C^{(1)}_{2,q}$ coefficient in Eq. (\ref{heavyz}) is related with the 
process $\gamma^*+h\gam \to h+G$, with a heavy quark in the initial state and 
therefore appears only in the ZVFNS approach. It is the only heavy-quark 
Wilson coefficient which we apply in the massless approximation.

The massless $C_{2,\gamma}^{(0)}(x)$ and $C_{2,G}^{(0)}(x)$ terms are not the 
only ones that must be subtracted. Still, we double count the corresponding 
collinear configurations which are a part of the DGLAP equations and manifest 
themselves via the splitting functions $P_{ij}$ and $k_i$. Such configurations
give rise to terms proportional to $\ln Q^2$ and through DGLAP equations are 
part of the heavy-quark densities $q_h\gam(x,Q^2)$. The same collinear 
configurations appear in massive heavy-quark Wilson coefficients 
$C^{h,(n)}_{2,j}$. Therefore, we must subtract the overlapping contributions. 
In the case of our model, where we neglect the higher order heavy-quark Wilson
coefficients, there are only two such terms, same as in the LO case described 
in \cite{cjkl} and \cite{cjk}. They are related to the 
$\gamma^* +\gamma \to h+\bar h$ and $\gamma^* +G \to h+\bar h$ processes. We
calculate the necessary subtraction terms from the approximated integration
of the corresponding parts of the DGLAP equations (\ref{DGLAP1}) and 
(\ref{DGLAP2}) over the region $(Q_0^2,Q^2)$ (see \cite{cjk}). The integration 
produces the $\ln\frac{Q^2}{Q_0^2}\cdot e_h^2\frac{\alpha}{2\pi} k_q^{(0)}(x)$ 
and $\ln\frac{Q^2}{Q_0^2}\cdot \frac{\alpha_s(Q^2)}{2\pi} 
\int_{\chi_h}^1\frac{dy}{y} P_{qG}^{(0)}(\frac{x}{y})G(y,Q^2)$ terms\footnote{
Notice that we neglected the $Q^2$ dependence of the strong coupling 
$\alpha_s$ and the gluon density $G(y,Q^2)$.}. These terms muliplied by $2x$ 
and $e_h^2$, according to the formula $F_2\gam(x,Q^2) = \sum_q xe_i^2 (q\gam + 
\bar q\gam)(x,Q^2)$, must be subtracted from the sum of Eq. (\ref{f2sum}).

\begin{figure}[t]
\begin{center}
\epsfxsize=20pc
\epsfbox{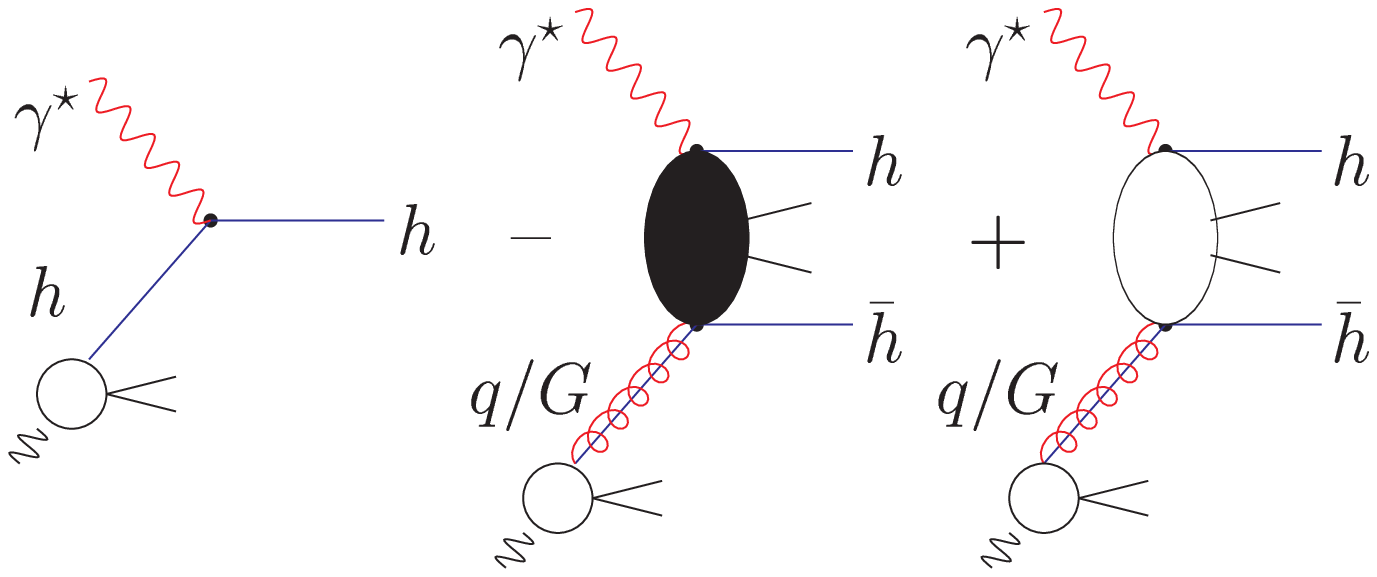}
\epsfxsize=15pc
\vskip -0.5cm
\epsfbox{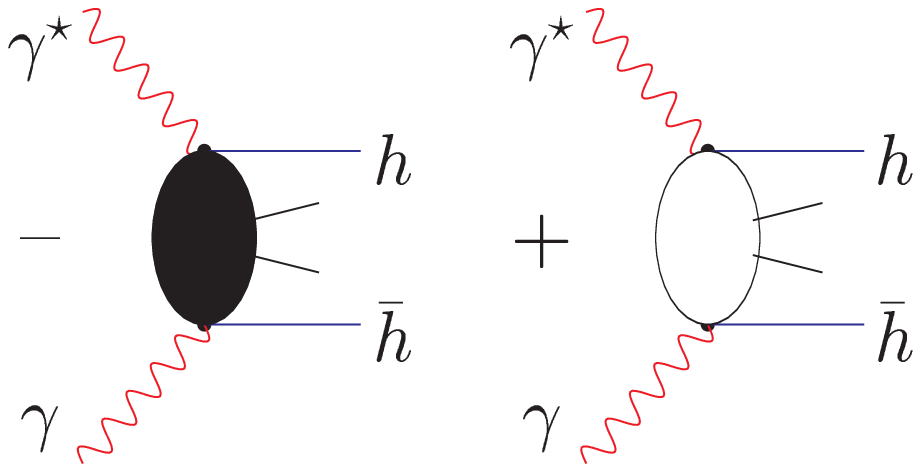}
\caption{\small Graphical representation of the contributions to the CJK NLO 
model. The first diagram represents the ZVFNS contribution. The third and 
fifth diagrams show the FFNS resolved-photon and direct-photon contributions, 
respectively. The second and forth diagrams are the corresponding subtraction 
terms.}
\label{tunggraph}
\end{center}
\end{figure}

The described summation and subtraction procedure leads to the following
formula for the \fun:
\ba
&& \frac{1}{x}F_2^{\gamma}(x,Q^2) = \sum_{i=1}^3 e_i^2 \left\{
(q_i\gam+\bar q_i\gam)(x,Q^2) + e_i^2 \frac{\alpha}{2\pi}C_{2,\gamma}^{(0)}(x)
 \right. \label{cjkf2} \\
&+& \left. \frac{\alsq}{2\pi}\int_x^1\frac{dy}{y}\left[
(q_i\gam+\bar q_i\gam)(y,Q^2)C^{(1)}_{2,q}(\frac{x}{y}) + 
G\gam(y,Q^2)C^{(1)}_{2,G}(\frac{x}{y}) \right] \right\} \nonumber \\
&+& \sum_{h(=c,b)}^2 e_h^2 \left\{ (q_h\gam+\bar q_h\gam)(x,Q^2) + 
e_h^2 \frac{\alpha}{2\pi} C_{2,\gamma}^{h,(0)}(x,\frac{Q^2}{m_h^2}) \right. 
\nonumber \\
&+& \frac{\alsq}{2\pi}\left[ \int_x^1\frac{dy}{y}
(q_h\gam+\bar q_h\gam)(y,Q^2)C^{(1)}_{2,q}(\frac{x}{y})+ 
\int_{\chi_h}^1\frac{dy}{y}
G\gam(y,Q^2)C_{2,G}^{h,(1)}(\frac{x}{y},\frac{Q^2}{m_h^2}) \right] \nonumber \\
&-& \left. \ln\frac{Q^2}{Q_0^2}\cdot 2e_h^2 \frac{\alpha}{2\pi} k_q^{(0)}(x) 
- \ln\frac{Q^2}{Q_0^2}\cdot 2\frac{\alsq}{2\pi} 
\int_x^1\frac{dy}{y} G\gam(y,Q^2)P_{qG}^{(0)}(\frac{x}{y}) \right\}. 
\nonumber
\ea
Its graphical representation is given in Fig. \ref{tunggraph}. If one 
includes the $O(\alpha_s^2)$ and $O(\alpha \alpha_s)$ heavy-quark 
contributions, the structure of the subtraction terms becomes more complicated,
since one has to subtract terms proportional to 
$k_q^{(1)}(x)$, $P_{qG}^{(1)}(x)$ and $P_{qq}^{(1)}(x)$. Moreover these 
functions are divergent at $x=1$ and therefore their integrals lead to 
numerical instabilities. Solution of that problem is beyond the scope of 
this work.

\subsection{The \ms and \dis factorization schemes\label{disacot}}

The introduction of the heavy-quark contributions to the \fun forces us to 
apply a modified \dis scheme in the CJK NLO model case. Let us recall that in 
this model apart from the massless, divergent at 
$x\to 1$ contribution of light-quarks, $C_{2,\gamma}^{(0)}(x)$, the massive,
finite $C_{2,\gamma}^{h,(0)}(x,\frac{Q^2}{m_h^2})$ term appears as the analogous 
heavy-quark contribution. In order to remove the large-$x$ divergence of the
$C_{2,\gamma}^{(0)}(x)$ we proceed in the same way as described in the FFNS 
approach, absorbing that term into the light-quark distribution functions 
$q_i\gam(x,Q^2)$. What about the massive 
$C_{2,\gamma}^{h,(0)}(x,\frac{Q^2}{m_h^2})$ 
contribution and the heavy-quark densities? The charm- and beauty-quark 
distributions in the CJK NLO model appear as a part of the massless ZVFNS
scheme. The evolution of all partons is performed through the common set of 
five-flavour DGLAP evolution equations. Therefore, the subtraction of the 
$C_{2,\gamma}^{(0)}(x)$ term from the Eq. (\ref{cjkf2}) and modification of the
$k_{q_i}$ and $k_G$ splitting functions (expressed in Eq. (\ref{modkqg}))
affect the heavy-quark densities in the same way as they affect the 
light-quark distributions. Namely, the following redefinition occurs:
\ba
q_h^{\dism}(x,Q^2) &=& q_h^{\msm}(x,Q^2) 
+ e_h^2 \frac{\alpha}{4\pi}C_{2,\gamma}^{(0),\msm}(x), \\
C_{2,\gamma}^{h,(0),\dism}(x,\frac{Q^2}{m_h^2}) &=& C_{2,\gamma}^{h,(0),\msm}(x,\frac{Q^2}{m_h^2}) 
+ \Delta C_{2,\gamma}^{(0)}(x). \nonumber
\ea
As $\Delta C_{2,\gamma}^{(0)}(x) = -C_{2,\gamma}^{(0),\msm}(x)$, see 
Eq. (\ref{delc2q}), the last equality takes form
\be
C_{2,\gamma}^{h,(0),\dism}(x,\frac{Q^2}{m_h^2}) = C_{2,\gamma}^{h,(0),\msm}(x,\frac{Q^2}{m_h^2}) -
C_{2,\gamma}^{(0),\msm}(x). \label{c2qdis}
\ee
In practice the subtraction of Eq. (\ref{c2qdis}) can not be performed in such
a simple way. Again, the $C_{2,\gamma}^{(0),\msm}(x)$ term on the right side 
would lead to a divergence at large $x$ and the advantage of the \dis scheme 
would be lost. We have to proceed in a different way.

We decide to resolve the above problem by loosing part of the information
brought by the massive $C_{2,\gamma}^{h,(0),\msm}(x)$ contribution. That term
calculated in the $m_h^2 \approx 0$ approximation leads to (see Appendix A)
\ba
C_{2,\gamma}^{h,(0),\msm}(x,\frac{Q^2}{m_h^2}) &=& C_{2,\gamma}^{(0),\msm}(x) 
+ e_h^4 2k_q^{(0),\msm}(x)\ln \frac{Q^2}{m_h^2} 
\label{c2qmsm} \\
&+& 6e_h^4\left[-\beta x(1-x)\frac{4m_h^2}{Q^2} +
\left(x(1-3x)\frac{4m_h^2}{Q^2}-x^2\frac{8m_h^4}{Q^4} \right)\ln 
\frac{1+\beta}{1-\beta} \right], \nonumber
\ea
where we keep the heavy-quark mass and 
$\beta=\sqrt{1-\frac{4m_h^2x}{(1-x)Q^2}}$ wherever it is possible.
Let us notice now that the term $\sim k_q^{(0),\msm}(x)$ of Eq. (\ref{c2qmsm}) 
differs from the first subtraction term in Eq. (\ref{cjkf2}) only by the
constant $\sim \ln \frac{m_h^2}{Q_0^2}$. We omit that constant and obtain in
the \ms scheme:
\ba
&&C_{2,\gamma}^{h,(0),\msm}(x,\frac{Q^2}{m_h^2}) -
\ln\frac{Q^2}{Q_0^2}\cdot 2e_h^4 k_q^{(0),\msm}(x)\approx \\
&\approx&  C_{2,\gamma}^{(0),\msm}(x) 
 + 6e_h^4\left[-\beta x(1-x)\frac{4m_h^2}{Q^2} +
\left(x(1-3x)\frac{4m_h^2}{Q^2}-x^2\frac{8m_h^4}{Q^4} \right)\ln 
\frac{1+\beta}{1-\beta} \right], \nonumber
\ea
which according to Eq. (\ref{c2qdis}) leads to the following relation in the
\dis scheme
\ba
&& \frac{\alpha}{2\pi} C_{2,\gamma}^{h,(0),\dism}(x,\frac{Q^2}{m_h^2}) -
\ln\frac{Q^2}{Q_0^2}\cdot 2e_h^4 \frac{\alpha}{2\pi} k_q^{(0),\msm}(x) \approx
\\
&\approx& 3e_h^4\frac{\alpha}{\pi} \left[-\beta x(1-x)\frac{4m_h^2}{Q^2} +
\left(x(1-3x)\frac{4m_h^2}{Q^2}-x^2\frac{8m_h^4}{Q^4} \right)\ln 
\frac{1+\beta}{1-\beta} \right]. \nonumber
\ea

In the following we use only the \dis factorization scheme, and omit the 
superscript for simplicity.


\subsection{The $\chi_h$ variables}

Let us recall that VFNS schemes widely utilize an approach in which 
$q_h\gam(x,Q^2)=0$ for all $Q^2$ values smaller than $m_h^2$ and 
$q_h\gam(x,Q^2)$ becomes nonzero for $Q^2 \ge m_h^2$. At each such step the 
number of active quarks grows by one. This approach leads to unphysical 
heavy-quark densities 
near their production thresholds. The kinematical threshold of the heavy-quark
production, and therefore its non-zero density in the photon, is given in 
DIS$_{e\gamma}$ by the total energy of the $e\gamma^*$ process, $W.$ As 
discussed in section 2 $W$ must be greater than the sum of two masses of the 
heavy quark in hand, $W>2m_h$. $W$ and $Q^2$ are connected, $W^2=(1-x)Q^2/x$, 
which is the reason why the ZVFNS condition $Q^2\ge m_h^2$ for treating the 
heavy quark $h$ as a parton is too simple.

In the ACOT($\chi$) scheme used by the CJK NLO model the number of active 
quarks is equal $N_f=5$ at the whole range of the DGLAP evolution. To ensure 
that a heavy-quark, $q_h\gam(x,Q^2)$, distribution disappears when $W \to 2m_h$
one introduces a new variable $\chi_h \equiv x(1+4m_h^2/Q^2)$ and calculates 
$q_h\gam(\chi_h,Q^2)$ in place of the $q_h\gam(x,Q^2)$. At the threshold, when 
$W=2m_h$, we obtain $\chi_h=1$ and the corresponding heavy-quark density 
vanishes as desired. The form of the $\chi_h$ variables could be chosen 
differently if only their above threshold behaviour was preserved. We chose to 
apply the same form as appears in the integration of the massive heavy-quark 
Wilson coefficients with the parton distributions, see also \cite{acot}. That 
way the same variable takes care of the proper vanishing of the $h$-quark 
density and of the 
$\sim \int_{\chi_h}^1\frac{dy}{y}G\gam(y,Q^2)C_{2,G}^{h,(1)}(\frac{x}{y},\frac{Q^2}{m_h^2})$
contribution to \fun at the kinematic limit. For the consistency of the 
approach also in the subtraction
$\sim \int_x^1\frac{dy}{y} P_{qG}^{(0)}(\frac{x}{y})G\gam(y,Q^2)$
term we exchange $x$ with $\chi_h$ and that way enforce its proper kinematic
behaviour.

As described in \cite{cjkl} and \cite{cjk} the subtraction $\sim k_q^{(0)}(x)$
term does not vanish as $W\to~2m_h$ even when one replaces $x$ with $\chi_h$. 
In the NLO case we avoid that problem because, as discussed in section 
\ref{disacot}, that contribution cancels in the modified \dis factorization 
scheme.

Unfortunately, a particular heavy-quark contribution appears in the NLO 
analysis which causes new problems at the threshold. The contribution
$\sim \int_x^1\frac{dy}{y}(q_h\gam+\bar q_h\gam)(y,Q^2)C^{(1)}_{2,q}(\frac{x}{y})$  
from the ZVFNS scheme could be rewritten in two ways, as $\sim 
\int_{\chi_h}^1\frac{dy}{y}(q_h\gam+\bar q_h\gam)(y,Q^2)C^{(1)}_{2,q}(\frac{x}{y})$ 
or as $\sim 
\int_{x}^1\frac{dy}{y}(q_h\gam+\bar q_h\gam)(\gamma_h,Q^2)C^{(1)}_{2,q}(\frac{x}{y})$,
with $\gamma_h$ defined analogically to $\chi_h$ $(\gamma_h=y(1+4m_h^2/Q^2))$.
The first approach leads to numerical instabilities for $y$ values close to 
$x$ originating from the $\frac{1+x^2}{1-x}\ln \frac{1-x}{x}$ term in 
$C^{(1)}_{2,q}(\frac{x}{y})$. Second form can not be adopted for the technical
reasons in the Mellin space. Therefore, we use the unmodified formula of 
Eq. (\ref{cjkf2}). Finally, to avoid the non-zero heavy-quark contributions in
the kinematically forbidden $(x,Q^2)$ region we impose by hand the following 
constraint: 
$\int_x^1\frac{dy}{y}(q_h+\bar q_h)(y,Q^2)C^{(1)}_{2,q}(\frac{x}{y}) = 0$
for $\chi_h>1$. Obviously, the same reasoning would apply to higher order 
$C^{(j)}_{2,q}$ coefficients neglected in the CJK NLO model.


\subsection{Formula for $F_2^{\gamma}$}

Summing the light- and heavy-quark contributions, converting them into the 
\dis factorization scheme and finally introducing the $\chi_h$ variables we 
obtain the final formula of the \fun in the CJK NLO model:
\ba
&& \frac{1}{x}F_2^{\gamma}(x,Q^2) = \label{f2cjk} \\
&& \sum_{i=1}^3 e_i^2 \left\{
(q_i\gam+\bar q_i\gam)(x,Q^2) + \frac{\alsq}{2\pi}\int_x^1\frac{dy}{y}\left[
(q_i\gam+\bar q_i\gam)(y,Q^2)C^{(1)}_{2,q}(\frac{x}{y}) + 
G\gam(y,Q^2)C^{(1)}_{2,G}(\frac{x}{y}) \right] \right\} \nonumber \\
&+& \sum_{h=1}^2 e_h^2 \left\{ (q_h\gam+\bar q_h\gam)(\chi_h,Q^2) 
+ \frac{\alsq}{2\pi} \left[ \int_x^1\frac{dy}{y}
(q_h\gam+\bar q_h\gam)(y,Q^2)C^{(1)}_{2,q}(\frac{x}{y}) \right. \right. \nonumber \\
&+& \left. \left.
\int_{\chi_h}^1\frac{dy}{y} 
G\gam(y,Q^2)C_{2,G}^{h,(1)}(\frac{x}{y},\frac{Q^2}{m_h^2}) 
- \ln\frac{Q^2}{Q_0^2}\cdot 2 
\int_{\chi_h}^1\frac{dy}{y} G\gam(y,Q^2)P_{qG}^{(0)}(\frac{\chi_h}{y})
\right] \right\} \nonumber \\
&+& 3e_h^2\frac{\alpha}{\pi} \left[-\beta x(1-x)\frac{4m_h^2}{Q^2} +
\left(x(1-3x)\frac{4m_h^2}{Q^2}-x^2\frac{8m_h^4}{Q^4} \right)\ln 
\frac{1+\beta}{1-\beta} \right]. \nonumber
\ea

To prevent the unphysical situation, in which the ZVFNS scheme contribution
to \fun is negative, or in other words in which the total \fun is smaller than 
the FFNS scheme contribution, as in \cite{cjkl,cjk,jank}, seperately for each 
heavy quark, we impose an additional positivity condition, here in form:
\be
(q_h\gam+\bar q_h\gam)(\chi_h,Q^2) 
+ \frac{\alpha_s}{2\pi} \left[ \int_x^1\frac{dy}{y}
(q_h\gam+\bar q_h\gam)(y,Q^2)C^{(1)}_{2,q}(\frac{x}{y}) - \ln\frac{Q^2}{Q_0^2}\cdot 2 
\int_{\chi_h}^1\frac{dy}{y} G\gam(y,Q^2) P_{qG}^{(0)}(\frac{\chi_h}{y})
\right] \geq 0.
\ee


\section{Global fits - solving the DGLAP evolution}

In this section the technical details of the solution of the DGLAP evoultion 
equations are shortly described.


\subsection{Mellin moments}

The DGLAP evolution equations (\ref{DGLAP1}) and (\ref{DGLAP2}) as well as the 
structure function formulae (\ref{f2ffns1}) and (\ref{f2cjk}) are very much 
simplified if they are transformed into the Mellin-moments space. The $n$-th 
moment, where $n$ is a complex number, for any function $f$, generally 
depending on $x$ and $Q^2$ is defined by 
$f^n(Q^2) = \int_0^1 x^{n-1}f(x,Q^2)dx$. We aplly this definition to the 
parton distributions, splitting functions and the light-quark Wilson 
coefficients $C_{2,i}^{(k)}(x)$. Their convolution integrals in the $(x,Q^2)$ 
space transform into multiplication of the corresponding moments in the Mellin
space. This simple correspondence does not hold in the case of the 
convolutions with the heavy-quark Wilson coefficient 
$C_{2,G}^{h,(1)}(x,\frac{Q^2}{m_h^2})$ and the $P_{qG}^{(0)}(x)$ splitting 
function because of the integration limit $\chi_h$. Therefore those 
integrations must be performed in the $(x,Q^2)$ space. All results calculated 
in the Mellin space are transformed through the inverse Mellin transformation 
to obtain their values in the $(x,Q^2)$ space.


\subsection{Nonsinglet and singlet parton densities}

In the photon case we define the singlet ($\Sigma\gam$) and non-singlet 
($f_{\mathrm{NS}(N_f)}\gam$) quark distributions as
\ba
f_{\mathrm{NS}(N_f)}\gam(x,Q^2) &=& \sum_{i=1}^{N_f} 
(e_i^2-\langle e^2 \rangle) 
\left[ q_i\gam(x,Q^2)+ \bar{q}_i\gam(x,Q^2) \right],\\
\Sigma\gam(x,Q^2) &=& \sum_{i=1}^{N_f}
[q_i^{\gamma}(x,Q^2)+\bar q_i^{\gamma}(x,Q^2)]
\ea
with
\be
\langle e^k \rangle = N_f^{-1}\sum_{i=1}^{N_f} e_i^k.
\ee
For the singlet density we sum over all quark flavours appearing in the 
evolution, the active quarks, $N_f=5$. Although in CJK model $N_f=5$, for the
non-singlet case it is necessary to calculate all, $q_{\mathrm{NS}(3)}\gam, 
q_{\mathrm{NS}(4)}\gam$ and $q_{\mathrm{NS}(5)}\gam(x,Q^2)$ to obtain all six 
parton distributions, see \cite{cjkl}.

The evolution of the non-singlet distributions is governed by the simplified 
DGLAP equation. In the Mellin moments space it reads
\be
\frac{df_{\mathrm{NS}(N_f)}\gamn(Q^2)}{d\ln Q^2}
= \frac{\alpha}{2\pi}k^n_{\mathrm{NS}(N_f)} + \frac{\alsq}{2\pi}
P_{qq}^{\mathrm{NS},n} f_{\mathrm{NS}(N_f)}\gamn(Q^2).
\ee

The singlet distribution evolution is coupled with the evolution of the gluon 
density
\ba
\frac{d\Sigma\gamn(Q^2)}{d\ln Q^2}
&=& \frac{\alpha}{2\pi}k^n_{\Sigma} + \frac{\alsq}{2\pi}
\Bigg[ P_{\Sigma \Sigma}^n \Sigma\gamn(Q^2) + P_{\Sigma G}^n G\gamn(Q^2) \Bigg]
, \label{DGLAPS1}\\
\frac{dG\gamn(Q^2)}{d\ln Q^2}
&=& \frac{\alpha}{2\pi}k^n_G + \frac{\alsq}{2\pi} 
\Bigg[ P_{G\Sigma}^n \Sigma\gamn(Q^2) + P_{GG}^n G\gamn(Q^2) \Bigg].
\label{DGLAPS2}
\ea

Thererefore, one has to solve the matrix equation
\be
\frac{d\vec f_{\mathrm{S}}\gamn(Q^2)}{d\ln Q^2}
= \frac{\alpha}{2\pi}\vec k^n_{\mathrm{S}} + \frac{\alsq}{2\pi}
\hat{P}^{\mathrm{S},n} f_{\mathrm{S}}\gamn(Q^2).
\ee
with
\be
\vec q_{\mathrm{S}}\gamn =  \left[ \begin{array}{c} \Sigma^{\gamma, n} \\ 
                              G^{\gamma, n} \end{array} \right], \quad
\vec k_{\mathrm{S}}^n = \left[ \begin{array}{c} k_{\Sigma}^n \\ k_G^n
                  \end{array} \right], \quad
\hat{P}^{\mathrm{S},n} = 
\left[ \begin{array}{cc} P_{\Sigma \Sigma}^n & P_{\Sigma G}^n \\
                         P_{G\Sigma }^n & P_{GG}^n  \end{array} \right].
\ee

The new point-like splitting functions are defined as
\ba
k_{\mathrm{NS}(N_f)}(x) &=& 2N_f \left( \langle e^4 \rangle-{\langle
    e^2 \rangle}^2 \right) \left[ k_q^{(0)} + \frac{\alsq}{2\pi}k_q^{(1)} 
\right](x), \\
k_{\Sigma}(x) &=& 2N_f \langle e^2 \rangle \left[ k_q^{(0)} + 
\frac{\alsq}{2\pi}k_q^{(1)} \right](x). \nonumber
\ea

The formulae for the above spliting functions $P_{ij}(x,Q^2)$ can be found for 
instance in \cite{Stirling}. Their Mellin moments were calculated in 
\cite{Floratos}.


\subsection{Point- and hadron-like parts}

The solution to the DGLAP equations can be divided into the so called 
point-like (pl) part, related to a special solution of the full inhomogenous
equations and hadron-like (had) part, being a general solution of the 
homogenous equations. Their sum gives the parton density in the photon, so 
we have
\ba
f_{\mathrm{NS}(N_f)}\gamn(Q^2) &=& f_{\mathrm{NS}(N_f),\had}\gamn(Q^2) + 
f_{\mathrm{NS}(N_f),\pl}\gamn(Q^2), \\
\vec f_{\mathrm{S}}\gamn(Q^2) &=& \vec f_{\mathrm{S},\had}\gamn(Q^2) + 
\vec f_{\mathrm{S},\pl}\gamn(Q^2).
\ea

The non-singlet hadron-like solution is given as
\be
q_{\mathrm{NS}(N_f),\had}\gamn(Q^2) = L^{-2P_{qq}^{(0),n}/\bz}
\left\{ 1 - \frac{\alsq-\alsm}{\pi\bz}\hat R  \right\}
q_{\mathrm{NS}(N_f),\had}\gamn(Q_0^2)
\label{hadsol}
\ee
and the non-singlet point-like solution is
\be
q_{\mathrm{NS}(N_f),\pl}\gamn(Q^2) = \frac{4\pi}{\alsq}
\left[1-L^{1-2P_{qq}^{(0)n}/\bz} \right] a^n + 
\left[1-L^{-2P_{qq}^{(0)n}/\bz} \right] b^n,
\label{plsol}
\ee
with
\be
a^n = \frac{\alpha}{2\pi\bz}\frac{k_{NS}^{(0)n}}{1-2P_{qq}^{(0)n}/\bz}, \quad
b^n = -\frac{1}{P_{qq}^{(0)n}} \left[ 2\hat R a^n + \frac{\alpha}{2\pi}\hat K
\right]
\ee
and
\be
L=\frac{\alsq}{\alsm}, \quad 
\hat R = P_{NS}^{(1),n}-\frac{\beta_1}{2\bz}P_{qq}^{(0),n}, \quad
\hat K = k_{NS}^{(1),n} -\frac{\beta_1}{2\bz}k_{NS}^{(0),n}.
\ee
Here $\beta_0$ and $\beta_1$ are the coefficients of the running strong 
coupling, $\alpha_s$, which we discuss in detail in section \ref{alfasec}.

The singlet point- and hadron-like solutions have the same form but are more
complicated than the corresponding non-singlet solutions. All formulae can be 
found for instance in \cite{vogt}.


\subsection{Input parton densities. VMD}

The hadron-like parts of the singlet and non-singlet densities, denoted in 
this section as $f_{\had}$, need input distributions. For this purpose we 
utilize the Vector Meson Dominance (VMD) model \cite{VMD}, where
\be
f_{\had}\gam(x,Q_0^2) = \sum_{V}\frac{4\pi \alpha}{\hat f^2_{V}}f^{V}(x,Q_0^2), 
\ee
with sum running over all light vector mesons (V) into which the photon can 
fluctuate. The parameters $\hat f^2_{V}$ can be extracted from the experimental
data on $\Gamma(V\to e^+e^-)$ width. In practice we take into account the  
$\rho^0$ meson while the contributions from the other mesons are accounted for 
via a parameter $\kappa$
\be
f_{\had}\gam(x,Q_0^2) = \kappa\frac{4\pi \alpha}{\hat f^2_{\rho}}f^{\rho}(x,Q_0^2),
\label{vmdfor}
\ee
which is left as a free parameter in the fits.

We use the input densities of the $\rho^0$ meson at low $Q_0^2$ in the form of
valence-like distributions both for the (light) quark ($v^{\rho}$) and gluon 
($G^{\rho}$) densities. All sea-quark distributions (denoted by $\zeta^{\rho}$)
are neglected at the input scale. At this scale, the densities 
$v^{\rho}, G^{\rho}$ and $\zeta^{\rho}$ are related, according to 
Eq. (\ref{vmdfor}) to the corresponding densities for a photon, see below.

The $v^{\rho}$ density is given by
\be
v^{\rho}(x,Q_0^2)= \frac{1}{4}
(u^{\rho^+}+\bar u^{\rho^-}+d^{\rho^-}+\bar d^{\rho^+})(x,Q_0^2),
\label{valence}
\ee
where from the isospin symmetry
\be
u^{\rho^+}(x,Q_0^2) = \bar u^{\rho^-}(x,Q_0^2)
= d^{\rho^-}(x,Q_0^2) = \bar d^{\rho^+}(x,Q_0^2).
\ee
Note that all the densities in Eq. (\ref{valence}) are normalized to 1,
eg. $\int^1_0 u^{\rho^+} dx =1$.

Two constraints should hold for the $v^{\rho}(x,Q_0^2)$ density, the first one 
is related to a number of valence quarks in the $\rho^0$ meson,
\be
\int_0^1 2 v^{\rho}(x,Q_0^2)dx = 2,
\label{const1}
\ee
the second constraint represents the energy-momentum sum rule:
\be
\int_0^1 x \left( 2v^{\rho}(x,Q_0^2)+G^{\rho}(x,Q_0^2) \right) dx = 1.
\label{const2}
\ee

We parametrize the input densities as follows:
\ba
x \zeta^{\rho}(x,Q_0^2)&=&0, \nonumber \\
xv^{\rho}(x,Q_0^2) &=& N_v x^{\alpha}(1-x)^{\beta}, \label{input1} \\
xG^{\rho}(x,Q_0^2) &=& \tilde N_G xv^{\rho}(x,Q_0^2)=
N_G x^{\alpha}(1-x)^{\beta}, \nonumber
\ea
where $N_G=\tilde N_GN_v$. Like in the LO analysis, \cite{cjkl,cjk,jank}, the 
input gluon distribution is proportional to the valence one and both have the 
valence-like form. Further, we impose two constraints given by Eqs. 
(\ref{const1}) and (\ref{const2}) in both types of models: FFNS$_{CJK}$ and CJK
NLO. These constraints allow us to express the normalization factors $N_{v}$ and
$N_G$ as functions of $\alpha, \beta$ and $\kappa$. This leaves these three 
parameters as the only free parameters to be fixed in the fits to the 
$F_2^{\gamma}$ experimental data.


\subsection{$\alpha_s$ running and values of $\Lambda^{(N_q)}$ \label{alfasec}}

We distinguish between the number of active quarks in the photon, $N_f$, and the
number of quarks contributing to the running of $\alpha_s$, which we denote by 
$N_q$, \cite{cjk}. The $N_f$ depends only on the choice of the type of the 
model: ones we decide to use the FFNS$_{CJK}$ or CJK NLO approach, then $N_f=3$ 
or 5, respectively, through the whole $Q^2$ range of evolution. On contrary 
$N_q$ depends on the $Q^2$ value. It is equal 3 when $Q^2<m_c^2$ and increases 
by one unit whenever $Q^2$ reaches a heavy-quark threshold, i.e. when 
$Q^2 = m_h^2$.

The running of the strong coupling constant in NLO is given then by the 
well-known formula:
\be
\frac{\alpha_s^{(N_q)}(Q^2)}{4\pi} = 
\frac{1}{ \beta_0 \ln (Q^2/\Lambda^{{(N_q)}^2}) } - \frac{\beta_1}{\beta_0^3}
\frac{\ln \ln(Q^2/\Lambda^{{(N_q)}^2}) }{(\Lambda^{{(N_q)}^2})^2}
\label{alphasold}
\ee
with
\be 
\beta_0 = 11-\frac{2}{3}N_q, \quad \beta_1 = 102-\frac{38}{3}N_q.
\ee

For $N_q=4$ we took the QCD scale $\Lambda^{(4)}$ equal to 280 MeV \cite{PDG}.
In order to ensure the continuity of the strong coupling constant at the 
heavy-quark thresholds the condition 
$\alpha_s^{(N_q)}(m_h^2)=\alpha_s^{(N_q+1)}(m_h^2)$ is imposed. 
We calculate values of the remaining $\Lambda^{(N_q)}$ constants according to 
the above condition. We utilizing the relation between constants 
$\Lambda^{(N_q)}$ and $\Lambda^{(N_q+1)}$ given in \cite{alphas}. That way we 
obtain $\Lambda^{(3)}=0.323$ MeV and $\Lambda^{(5)}=0.200$ MeV.

The parton evolution equations depend on $N_q$ through their dependence on 
$\alpha_s(Q^2)$ and independently $\beta_0$ and $\beta_1$, see Eqs. 
(\ref{hadsol}) and (\ref{plsol}). Therefore, because of the implicit 
introduction of the heavy-quark thresholds into the $\alpha_s$ running we must 
proceed in three steps to perform the DGLAP evolution. In the first step, 
describing the evolution from the input scale $Q_0$ to the charm-quark mass 
$m_c$, the hadronic input $q_{\mathrm{had}}\gam(x,Q_0^2)$ is taken from the VMD 
model. In the second step we evolve the parton distributions from $m_c$ to the 
beauty-quark mass, $m_b$. Then, a new hadronic input is given by the sum of 
the already evolved hadronic and point-like contributions. The point-like 
distribution at $Q^2=m_c^2$ becomes zero again. The same is repeated for 
$Q^2>m_b^2$.


\section{Global fits - results}

We have performed a series of fits to the \fun data \cite{CELLO}-\cite{HQ2}. 
All together 192 data points were used, including the recent results of the 
ALEPH Collaboration \cite{ALEPHnew}, which replaced the old prelimiary data 
\cite{ALEPHold}, as well as recent preliminary results of the DELPHI 
Collaboration \cite{DELPHInew} replacing the old prelimiary data 
\cite{DELPHIold}. In that case only the LEP1 data were applied because no 
final results for the LEP2 data are given in a form useful for our global 
fits\footnote{All LEP2 data are given in three sets obtained with the TWOGAM 
\cite{TWOGAM}, PHOJET \cite{PHOJET} and PYTHIA \cite{PYTHIA} Monte Carlo 
generators.} in \cite{DELPHInew}. Our fits based on the least square principle
(minimum of $\chi^2$) were done using \textsc{Minuit} \cite{minuit}. 
Systematic and statistical errors of data points were added in quadrature.

First, we made two test fits in which, following the analysis of the GRS group,
\cite{grs}, we assumed the input scale value $Q_0^2=0.4$ GeV$^2$. The results 
of both fits are presented in table \ref{tparam04}. The first two columns show
the quality of the fits, i.e. the total $\chi^2$ for 192 points and the 
$\chi^2$ per degree of freedom. The fitted values for parameters $\alpha$, 
$\beta$ and $\kappa$ are presented in the middle of the table with the 
symmetric parabolic errors obtained from \textsc{Migrad} requiring 
$\Delta \chi^2 = 1$\footnote{For test fits we applied \textsc{Migrad} instead 
of \textsc{Minos} which requires much more computer time}. In addition, the 
values for $N_v$ and $\tilde N_{g}$ obtained from these parameters, using the 
constraints (\ref{const1}) and (\ref{const2}), are given in the last two 
columns. As can be seen the fits give high $\chi^2$ per degree of freedom. 
Further, the relatively small $\kappa$ value indicates respectively small 
contribution of heavier mesons which is compensated by a very high, especially
in case of the CJK NLO model, valence-quark input ($N_v=9.26$).

{\small
\begin{table}[h]
\begin{center}
\renewcommand{\arraystretch}{1.5}
\begin{tabular}{|c|@{} p{0.1cm} @{}|c|c|@{} p{0.1cm} @{}|c|c|c|@{} p{0.1cm} @{}|c|c|}
\hline
 NLO models         && $\chi^2$ & $\chi^2/_{DOF}$ && $\kappa$ & $\alpha$ & $\beta$ && $N_v$ & $\tilde N_G$ \\
\hline
\hline
 FFNS$_{CJK}$1 && 318.0 & 1.68 && 1.40$\pm 0.06$ & 1.15$\pm 0.16$ & 1.24$\pm 0.42$ && 2.80 & 0.949 \\
\hline
 CJK           && 299.9 & 1.51 && 1.40$\pm 0.05$ & 1.38$\pm 0.18$ & 3.40$\pm 0.77$ && 9.26 & 2.19 \\
\hline
\end{tabular}
\caption{\small The $\chi^2$ and parameters of the test fits for 192 data 
points for FFNS$_{CJK}$1 NLO and CJK NLO models with assumed $Q_0^2 = 0.4$ 
GeV$^2$. The $\alpha$, $\beta$ and $\kappa$ symmetric parabolic errors obtained
from \textsc{Migrad} requiring $\Delta \chi^2 = 1$.}
\label{tparam04}
\end{center}
\end{table}}

In order to obtain better agreement with the data two other fits were 
performed, in which $Q_0^2$ was treated at first as a free parameter. For both
models we obtained very similar $Q_0^2 \approx 0.765$ GeV$^2$ 
($Q_0=0.875$ GeV) values. Therefore in our final CJK and FFNS$_{CJK}$NLO fits 
we fixed the input scale as $Q_0^2=0.765$ GeV$^2$. The $Q_0=0.875$ GeV value 
agrees with the $Q_0=0.85\pm 0.09$ GeV obtained in the recent NLO fit 
presented in \cite{klasen}. The results of both CJK and FFNS$_{CJK}$1 NLO fits
are presented in table \ref{tparam}. The errors shown in the table were 
obtained from \textsc{Minos} requiring $\Delta \chi^2 = 1$.

As expected, estimation of the $Q_0^2$ value through the test fits, allowed us 
to find sets of parameters which give better agreement of both models with the
data. We use these sets as the final results of our analysis. Comparison of 
the $\chi^2/_{DOF}$ values obtained in the test and final fits shows the 
$\sim 0.40$ and $\sim 0.30$ improvment in the FFNS$_{CJK}$1 and CJK NLO models
respectively. Further, in both cases we observe an increase of the $\kappa$ 
parameter with the simultaneous decrease of $\alpha, \beta$ and $N_v$. That 
change can be explained by the fact that in higher $Q^2$ valence distribution
broadens (change of $\alpha, \beta$) but at the same time the constraint 
(\ref{const1}) maintains its integral constant. Finally, the $\tilde N_G$
parameter grows at higher $Q_0^2$ because the gluon contribution to the 
photon structure increases with the increasing $Q^2$.

\begin{figure}
\hskip -1cm
\includegraphics[scale=1.0]{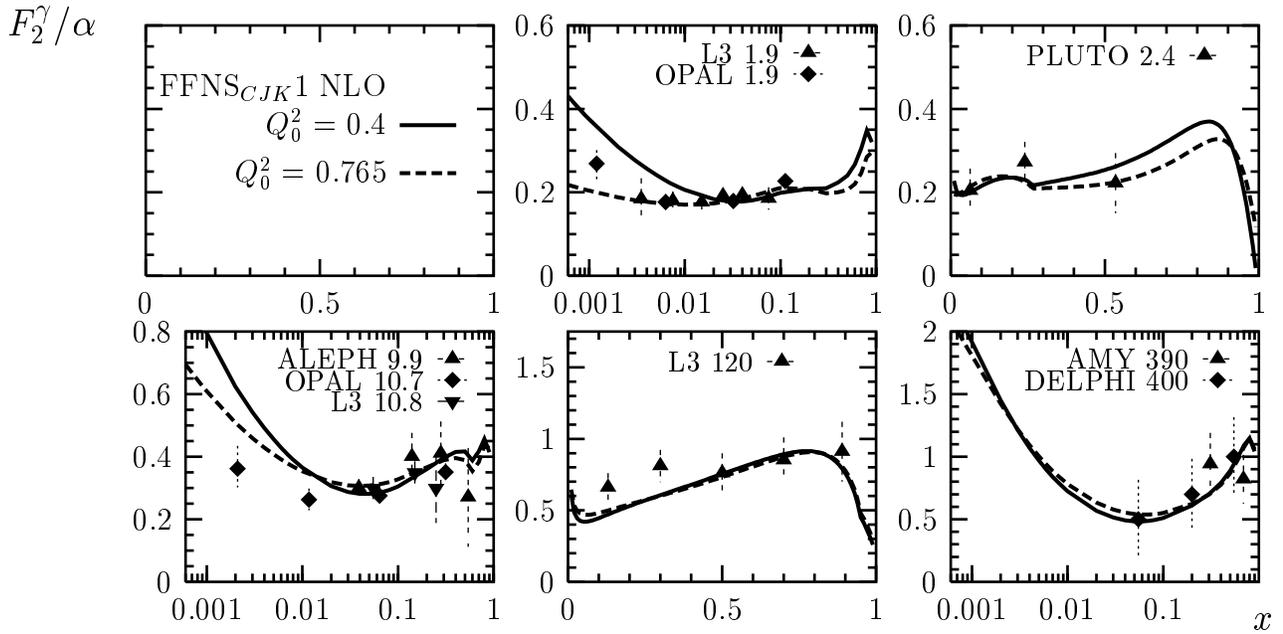}%
\caption{Predictions for the $F_2^{\gamma}(x,Q^2)/\alpha$ for the FFNS$_{CJK}$1
NLO model obtained by the main fit with $Q_0^2=0.765$ GeV$^2$ and by the
test fit with Q$_0^2=0.4$ GeV$^2$.}
\label{fitff04}
\end{figure}

\begin{figure}
\hskip -1cm
\includegraphics[scale=1.0]{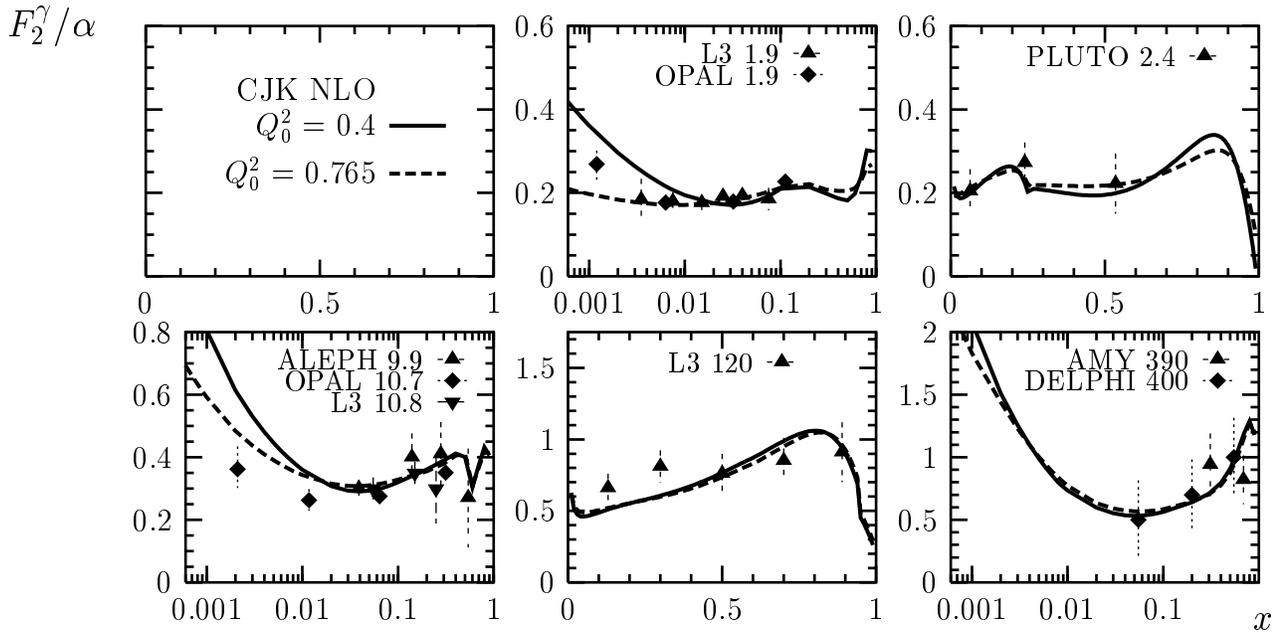}%
\caption{The same as in Fig. \ref{fitff04} for the CJK NLO model.}
\label{fitvf04}
\end{figure}

The choice of the \fun distributions predicted by the test and final fits are 
presented in Figs. \ref{fitff04} and \ref{fitvf04} for the FFNS$_{CJK}$1 and 
CJK NLO models, respectively. As the main difference we observe faster growth 
of the $F_2\gam$ function at small $x$ in the case of the test fits with 
$Q_0^2=0.4$ GeV$^2$, especially at small $Q^2$. The medium and high-$x$ 
predictions for the photon structure function are similar apart from the low 
$Q^2$ region.

\begin{table}[h]
\begin{center}
\renewcommand{\arraystretch}{1.5}
\begin{tabular}{|c|@{} p{0.1cm} @{}|c|c|@{} p{0.1cm} @{}|c|c|c|@{} p{0.1cm} @{}|c|c|}
\hline
 NLO models         && $\chi^2$ & $\chi^2/_{DOF}$ && $\kappa$ & $\alpha$ & $\beta$ && $N_v$ & $\tilde N_G$ \\
\hline
\hline
 FFNS$_{CJK}$1 && 243.3 & 1.29 && 2.288$^{+0.108}_{-0.096}$ & 0.502$^{+0.071}_{-0.066}$ & 0.690$^{+0.282}_{-0.252}$ && 0.685 & 2.369 \\
\hline
 CJK           && 256.8 & 1.37 && 2.662$^{+0.108}_{-0.099}$ & 0.496$^{+0.063}_{-0.057}$ & 1.013$^{+0.284}_{-0.255}$ && 0.745 & 3.056 \\
\hline
\end{tabular}
\caption{\small The $\chi^2$ and parameters of the final fits for 192 data 
points for FFNS$_{CJK}$1 NLO and CJK NLO models with assumed $Q_0^2 = 0.765$ 
GeV$^2$. The $\alpha$, $\beta$ and $\kappa$ errors obtained from \textsc{Minos}
requiring $\Delta \chi^2 = 1$.}
\label{tparam}
\end{center}
\end{table}

Moreover, we see that unlike in the LO case (see \cite{cjkl} and \cite{cjk}), 
and in the case of the test fits with $Q_0^2=0.4$ GeV$^2$, the obtained 
$\chi^2$ per degree of freedom is better in the standard type FFNS model than 
in the CJK model. The origin of the relatively high $\chi^2/_{DOF}$ in both 
fits comparing to much lower $\chi^2\approx 0.9$ obtained in \cite{klasen} is 
the same as in the LO case, discussed in detail in \cite{jank}. Namely, the
exclusion of the CELLO \cite{CELLO} and recent DELPHI \cite{DELPHInew} data 
from the fits leads to much better agreement among fits and the data.

\bigskip

Further, we performed test fits with the modified set of the input densities
(\ref{input1}). Namely, we used the input gluon distribution independent of the
valence one
\ba
x \zeta^{\rho}(x,Q_0^2)&=&0, \nonumber \\
xv^{\rho}(x,Q_0^2) &=& N_v x^{\alpha}(1-x)^{\beta}, \label{input2} \\
xG^{\rho}(x,Q_0^2) &=& N_G x^{\alpha_G}(1-x)^{\beta_G}. \nonumber
\ea
Fits gave very high values of the $\alpha_G$ and $\beta_G$ parameters, around 
14 and 16 respectively, leading to fast oscillations of the input gluon 
densities and all parton densities at low $Q^2$. The $N_G$ parameter calculated
using constraint (\ref{const2}) is of the order $10^9$. Moreover, the 
$\alpha_G$ and $\beta_G$ values have very high uncertainties and therefore it 
is hard to obtain the final convergence of the fits. Taking all this and the 
fact that the $\chi^2/_{DOF}$ values obtained in the test fits are slightly 
higher than the once presented above, we conclude that the choice of the simple
form of the input distributions of Eqs. (\ref{input1}) is very reasonable even 
at $Q_0^2\approx 0.8$ GeV$^2$.


\subsection{Comparison of the CJK and FFNS$_{CJK}$ NLO fits with $F_2\gam$ data and other NLO parametrizations}

\begin{figure}
\hskip -0.5cm
\includegraphics[scale=1.0]{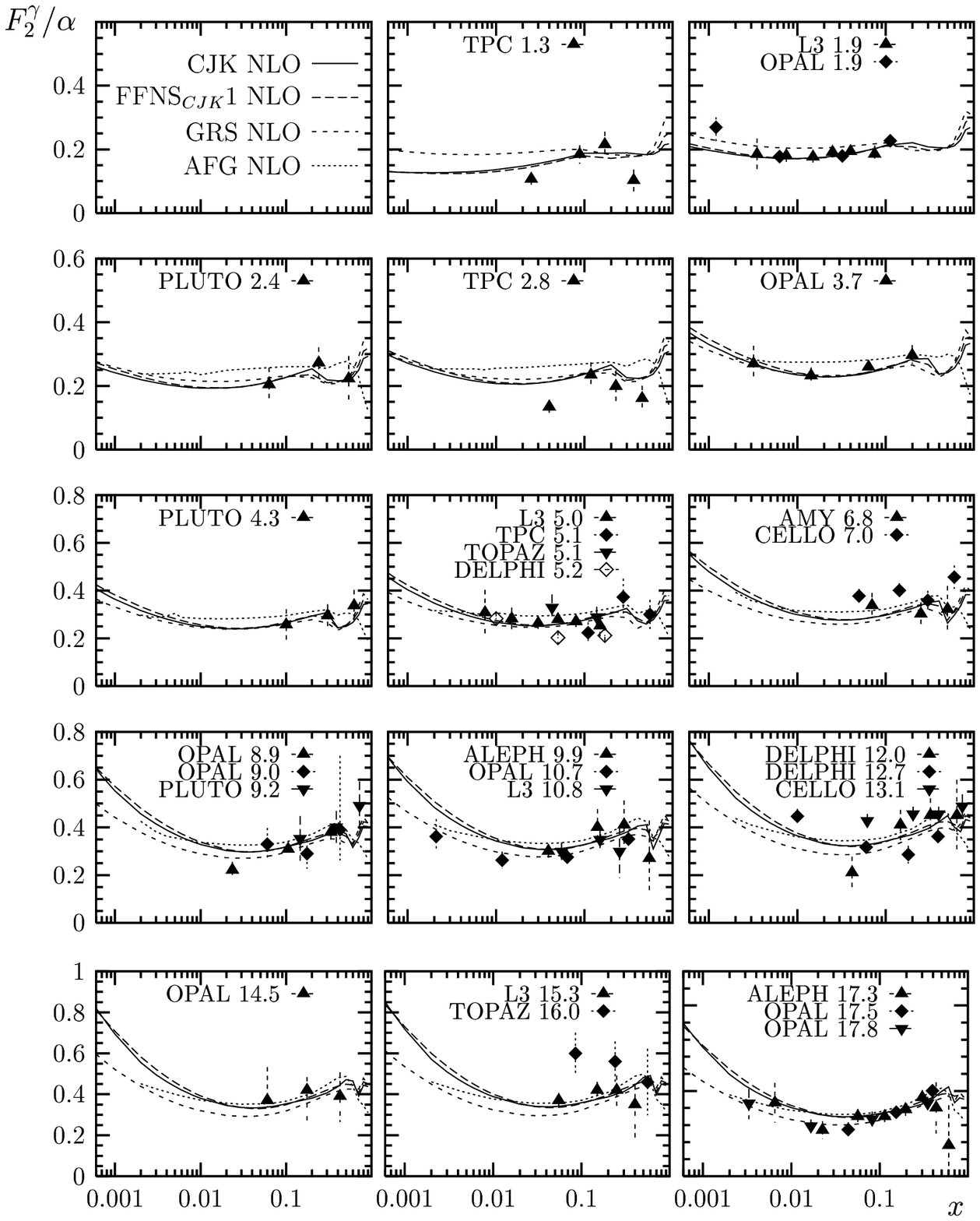}%
\vskip -0.2cm
\caption{Predictions for the $F_2^{\gamma}(x,Q^2)/\alpha$ for the CJK NLO
and FFNS$_{CJK}$ NLO models and GRS NLO \cite{grs} and AFG NLO \cite{afg}
compared with the experimental data \cite{CELLO}--\cite{HQ2}, for small and 
medium $Q^2$ as a function of $x$ (logarithmic scale). If a few values of 
$Q^2$ are displayed in the panel, the average of the smallest and biggest 
$Q^2$ was taken in the computation.}
\label{fitparam1}
\end{figure}

\begin{figure}
\hskip -0.5cm
\includegraphics[scale=1.0]{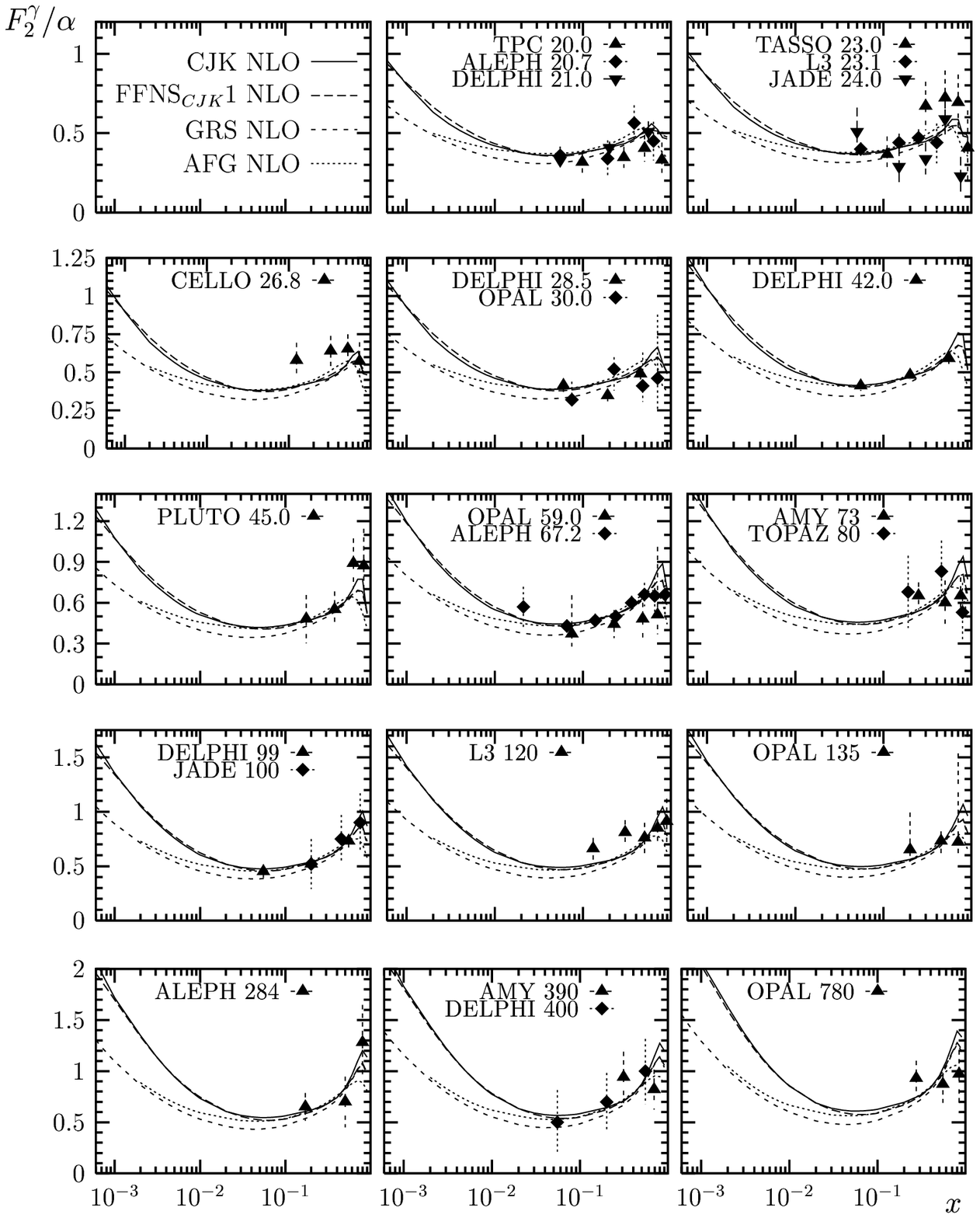}%
\caption{The same as in Fig. \ref{fitparam1}, for 
$Q^2 \geq 20 \mathrm{GeV}^2$.}
\label{fitparam2}
\end{figure}

\begin{figure}
\hskip -0.5cm
\includegraphics[scale=1.0]{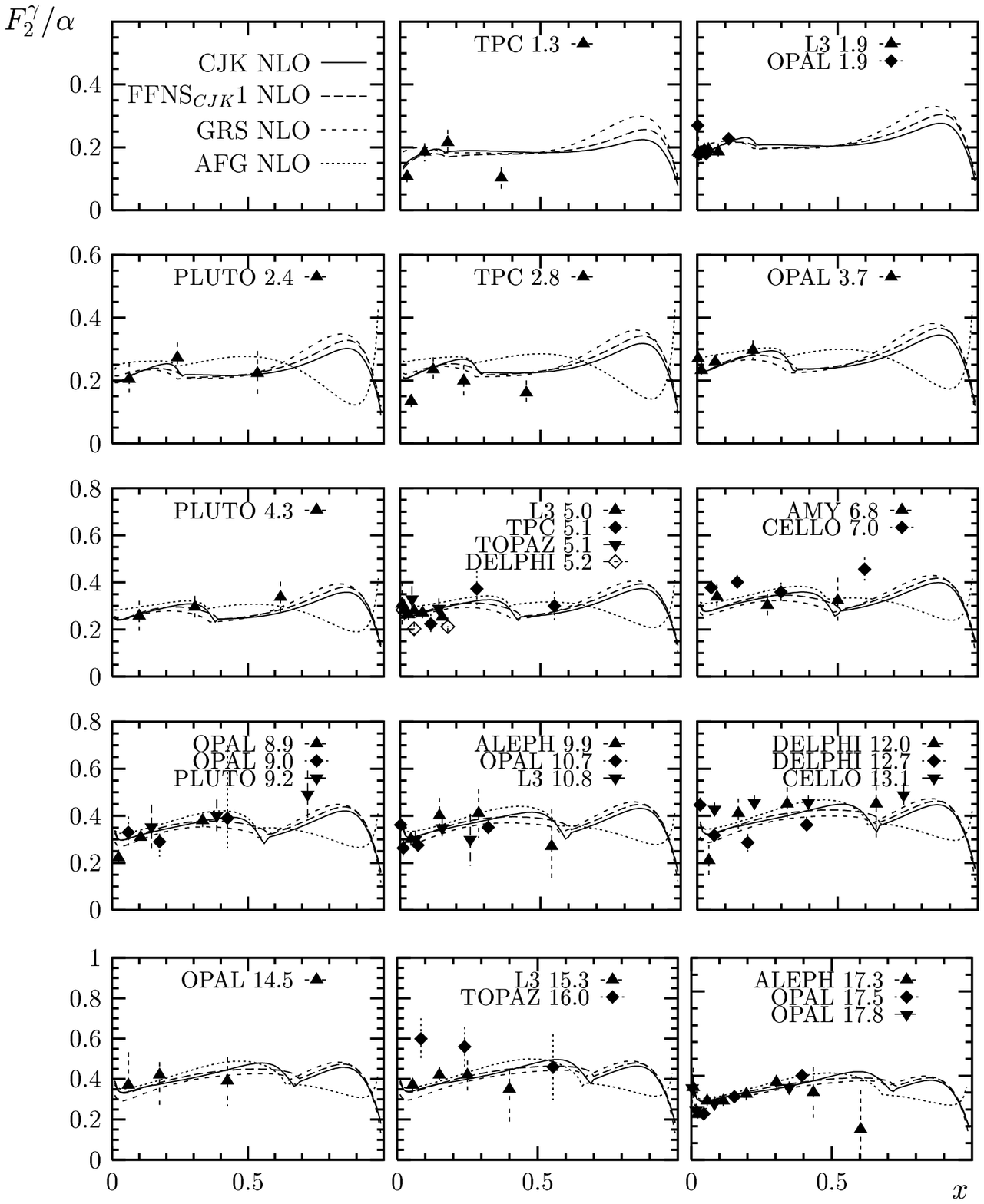}%
\caption{The same as in Fig. \ref{fitparam1} for a linear scale in $x$.}
\label{fitparam3}
\end{figure}

\begin{figure}
\hskip -0.5cm
\includegraphics[scale=1.0]{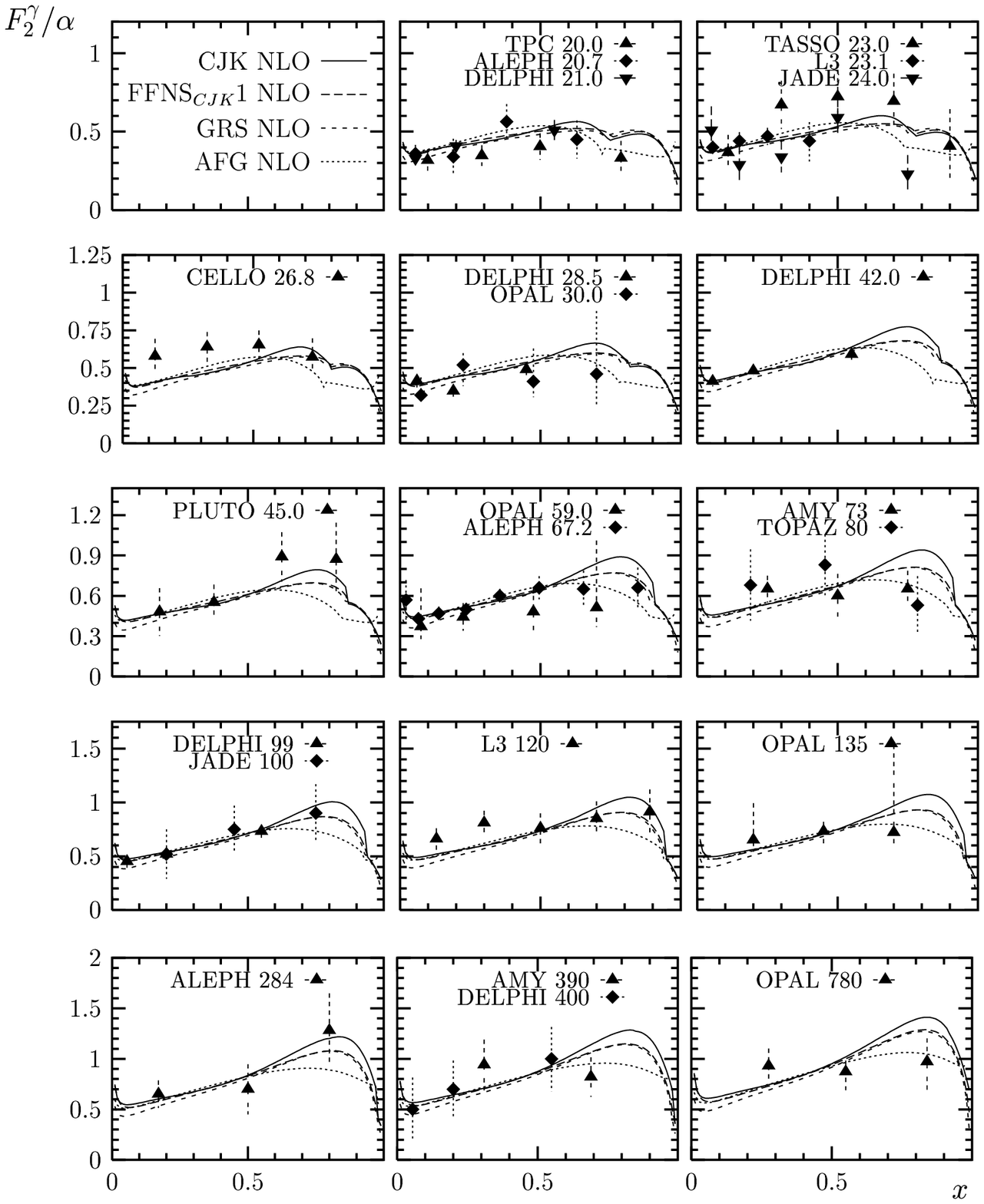}%
\caption{The same as in Fig. \ref{fitparam2} for a linear scale in $x$.}
\label{fitparam4}
\end{figure}

Figures \ref{fitparam1}--\ref{fitparam4} show a comparison of the CJK and 
FFNS$_{CJK}1$ NLO fits to \fun with the experimental data as a function of 
$x$, for different values of $Q^2$. Also a comparison with the GRS NLO 
\cite{grs} and AFG NLO \cite{afg} parametrizations is shown. (If a few values 
of $Q^2$ are displayed in a panel, the average of the smallest and biggest one
was taken in the computation.) As can be seen in Figs. \ref{fitparam1} and 
\ref{fitparam2}, both CJK and FFNS$_{CJK}1$ NLO models predict a much steeper 
behaviour of the \fun at small $x$ with respect to other parametrizations. 
In the region of medium and high $x$, the behaviour of the \fun obtained from 
the FFNS$_{CJK}1$ and CJK NLO fits is similar to the one computed with the 
GRS NLO parametrization while it is very different to the AFG parametrization
prediction, see Figs. \ref{fitparam3} and \ref{fitparam4}. Below the 
charm-quark threshold the CJK model gives the highest $F_2\gam$ and the GRS 
NLO parametrization the lowest $F_2\gam$ of the three similar predictions 
while above the threshold that order is inverted. Finally, at high $Q^2$ and 
high $x$ the $F_2\gam$ computed in the FFNS$_{CJK}1$ NLO model and GRS NLO 
parametrization become undistinguishable while the CJK NLO prediction is 
visibly larger.

\begin{figure}
\hskip -0.5cm
\includegraphics[scale=1.0]{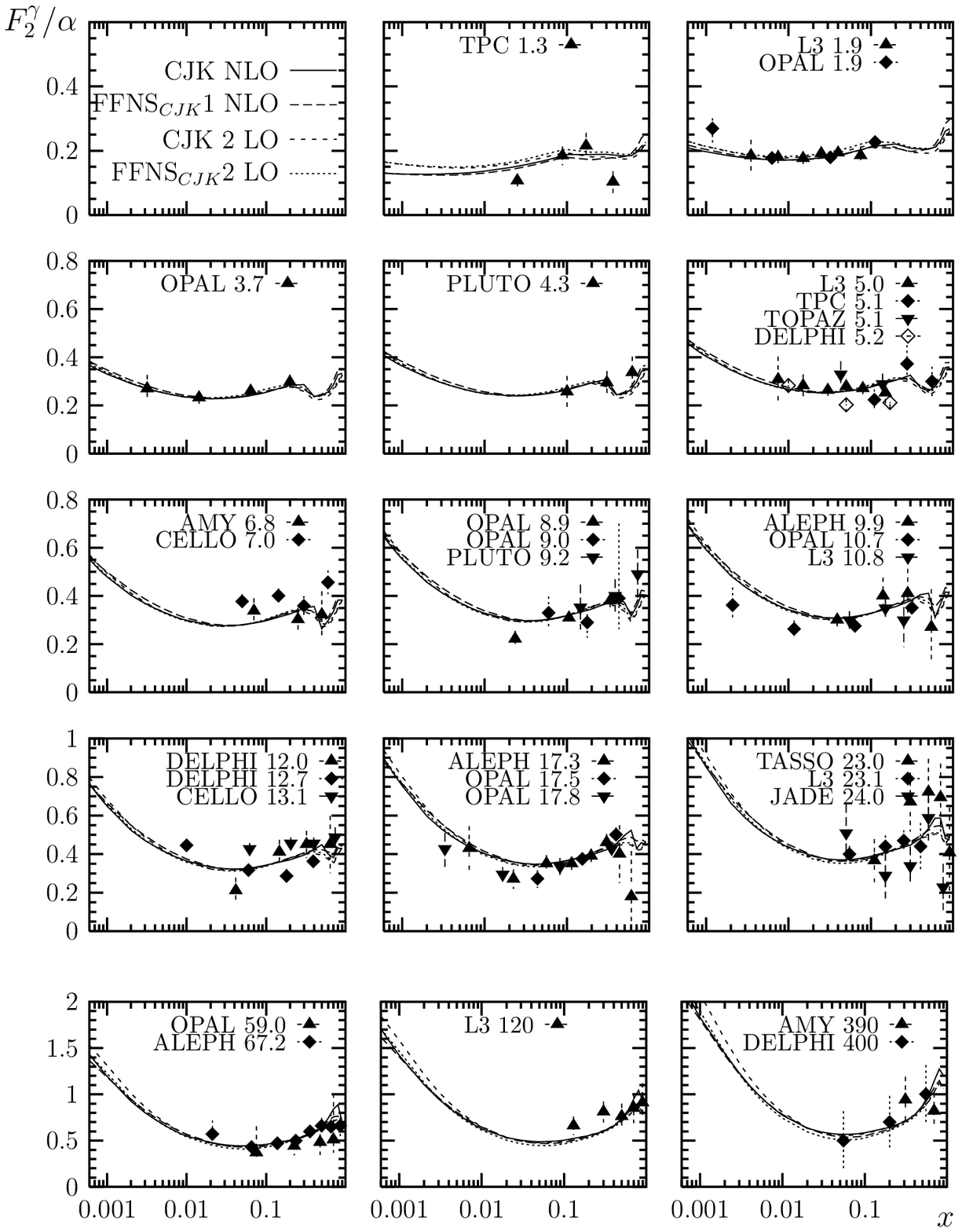}%
\vskip -0.2cm
\caption{Predictions for the $F_2^{\gamma}(x,Q^2)/\alpha$ for the CJK NLO
and FFNS$_{CJK}$ NLO models compared with corresponding LO fits from \cite{cjk}
and with the experimental data \cite{CELLO}--\cite{HQ2}, for small and 
medium $Q^2$ as a function of $x$ (logarithmic scale). If a few values of 
$Q^2$ are displayed in the panel, the average of the smallest and biggest 
$Q^2$ was taken in the computation.}
\label{fitloho1}
\end{figure}

\begin{figure}
\hskip -0.5cm
\includegraphics[scale=1.0]{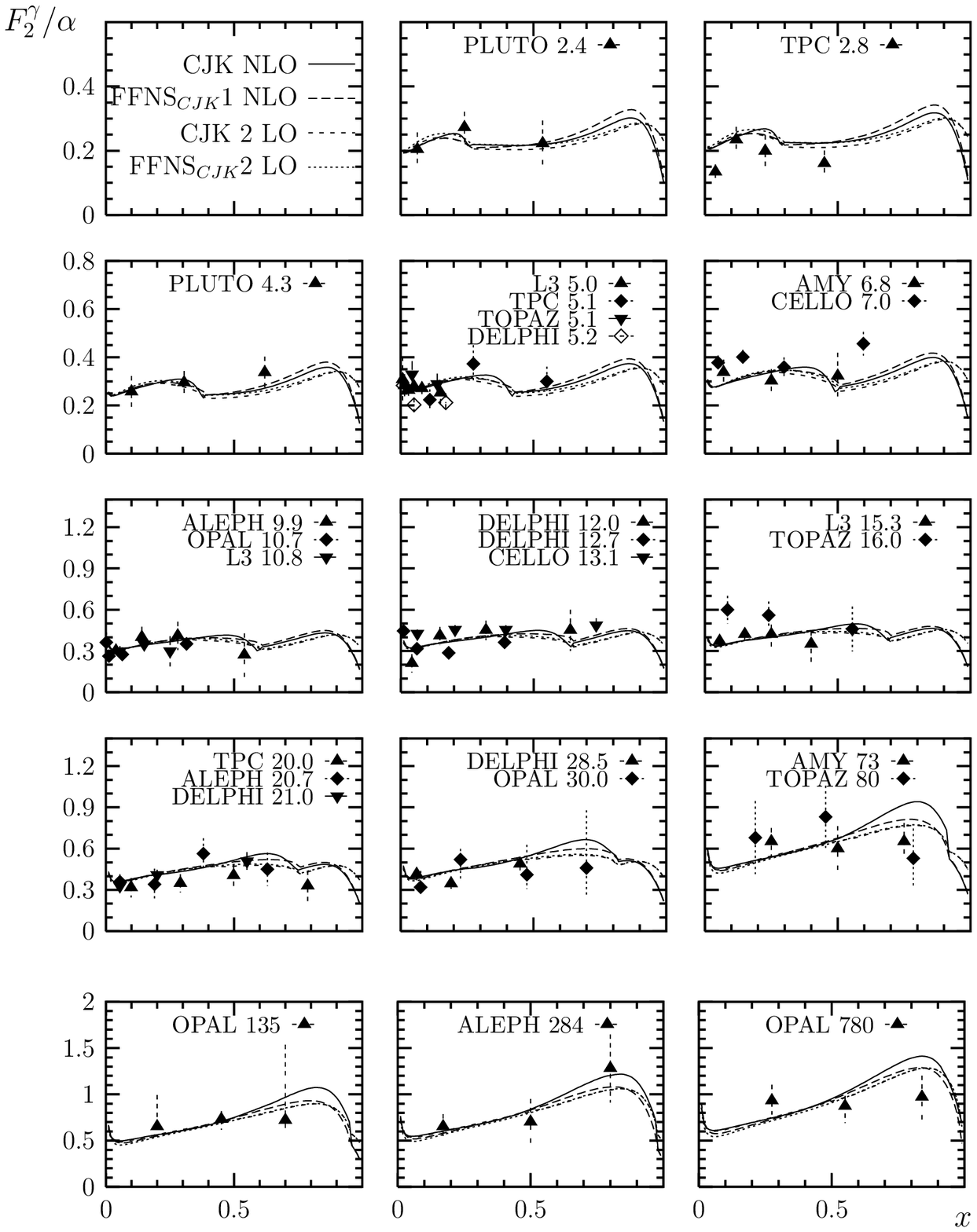}%
\caption{The same as in Fig. \ref{fitloho1} for a linear scale in $x$.}
\label{fitloho2}
\end{figure}

Next, in Figs. \ref{fitloho1} and \ref{fitloho2} we compare the CJK and 
FFNS$_{CJK}1$ NLO fits with the corresponding CJK and FFNS$_{CJK}2$ LO fits 
presented in \cite{cjk}. Figure \ref{fitloho1} shows very small differences,
only in the low $x$ region the FFNS$_{CJK}1$ NLO model gives greater values 
then other models and parametrizations. On the other hand, as it can be seen in
Fig. \ref{fitloho2}, the NLO models predict higher $F_2\gam$ than the LO ones
at the $x$ region right below the charm-quark threshold. Moreover the NLO 
predictions vanish more rapidly at $x\to 1$.

\begin{figure}
\includegraphics[scale=1.0]{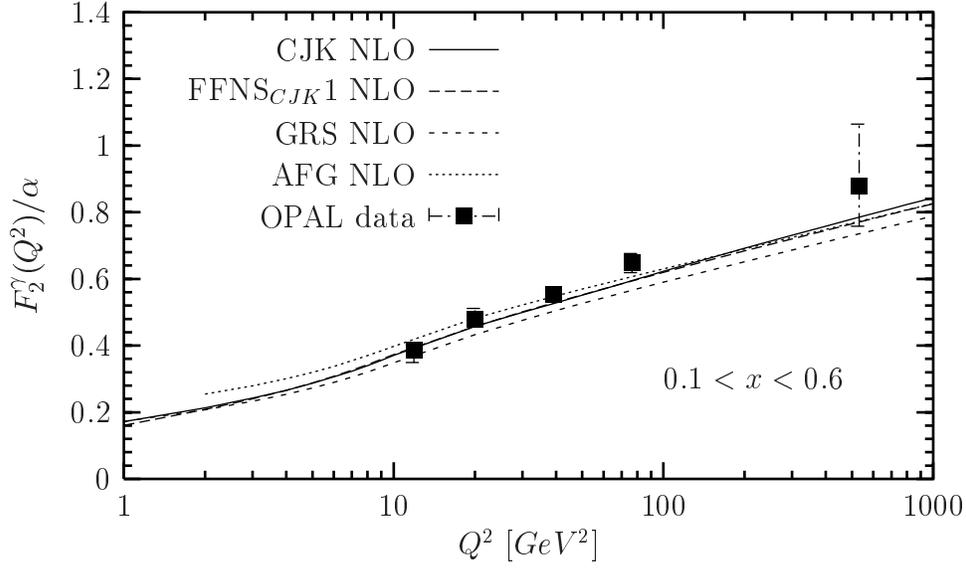}%
\caption{Comparison of the OPAL data \cite{HQ2} for the
$Q^2$-dependence of the averaged over $0.1<x<0.6$ $F_2^{\gamma}/\alpha $
with the predictions of the CJK NLO and FFNS$_{CJK}$ NLO models.}
\label{evol1}
\end{figure}

\begin{figure}[h]
\includegraphics[scale=1.0]{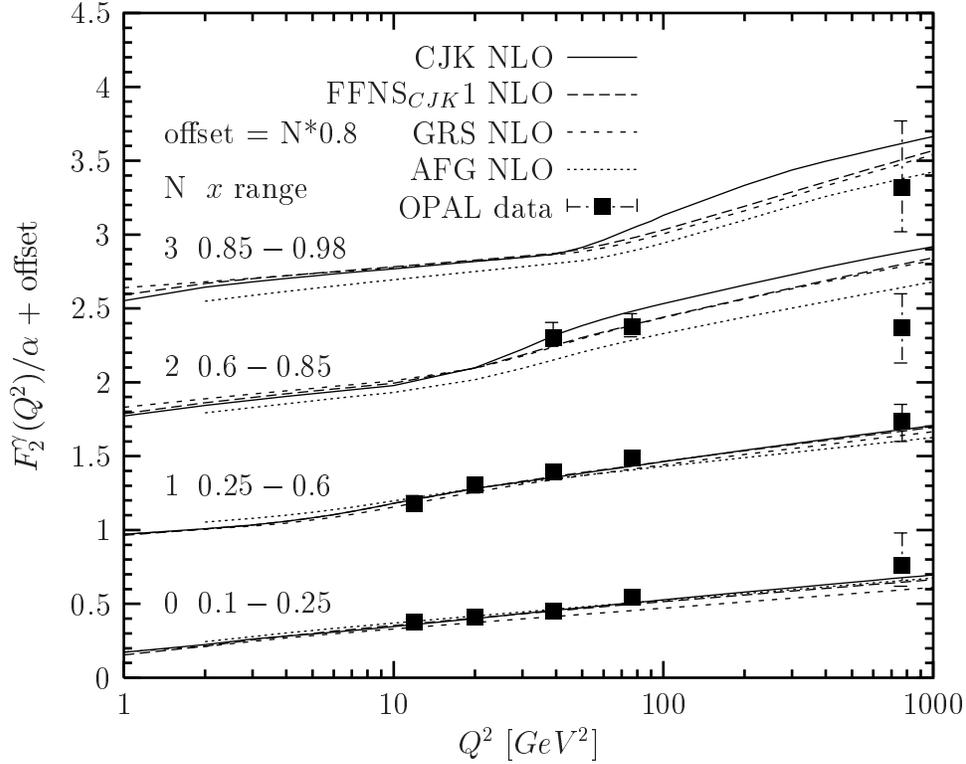}%
\caption{
As in Fig.~\ref{evol1} for \fund/$\alpha$, averaged over four different $x$ 
ranges.}
\label{evol2}
\end{figure}

Apart from the direct comparison with the photon structure-function data,
we perform another comparison, this time with LEP data that were not used 
directly in our analysis. Figures \ref{evol1} and \ref{evol2} present the 
predictions for \fund, averaged over various $x$ regions, compared with the 
recent OPAL data \cite{HQ2}. For comparison, the results from the GRS and 
AFG NLO parametrizations are shown as well. In Fig. \ref{evol1} we observe 
that in the medium-$x$ range, $0.1<x<0.6$, all models give expectations of the 
same shape, all in fairly good agreement with the experimental data. Still, in
small $Q^2$ the $F_2\gam$ computed with the AFG NLO parametrization is above 
the others, and starting at approximately $Q^2=3$ GeV$^2$ the predictions of 
the GRS NLO parametrization are lower than the other three. In the $x$ ranges
presented in Fig. \ref{evol2} we also notice many differences among the 
averaged \fun predictions computed in various models. The only exception give 
the FFNS$_{CJK}1$ NLO and the GRS NLO predictions which are very close to each
other in the whole $Q^2$ range and in all, apart from the lowest, $x$ ranges 
examined. Let us now name all the dissimilarities observed in Fig.\ref{evol2}.
Firstly, the $F_2\gam(Q^2)$ predicted by the CJK NLO model shows clear change 
of the behaviour in the high $Q^2$ and high $x$ range where we observe its
larger increase than predicted by the other models. Secondly, the expectations
of the AFG parametrization in the same regions (also for small $Q^2$) lie much 
below the other results. Finally, the GRS NLO parametrization predicts 
slightly lower $F_2\gam(Q^2)$ values than the remaining models.


\subsection{Parton densities}

\begin{figure}
\includegraphics[scale=1.0]{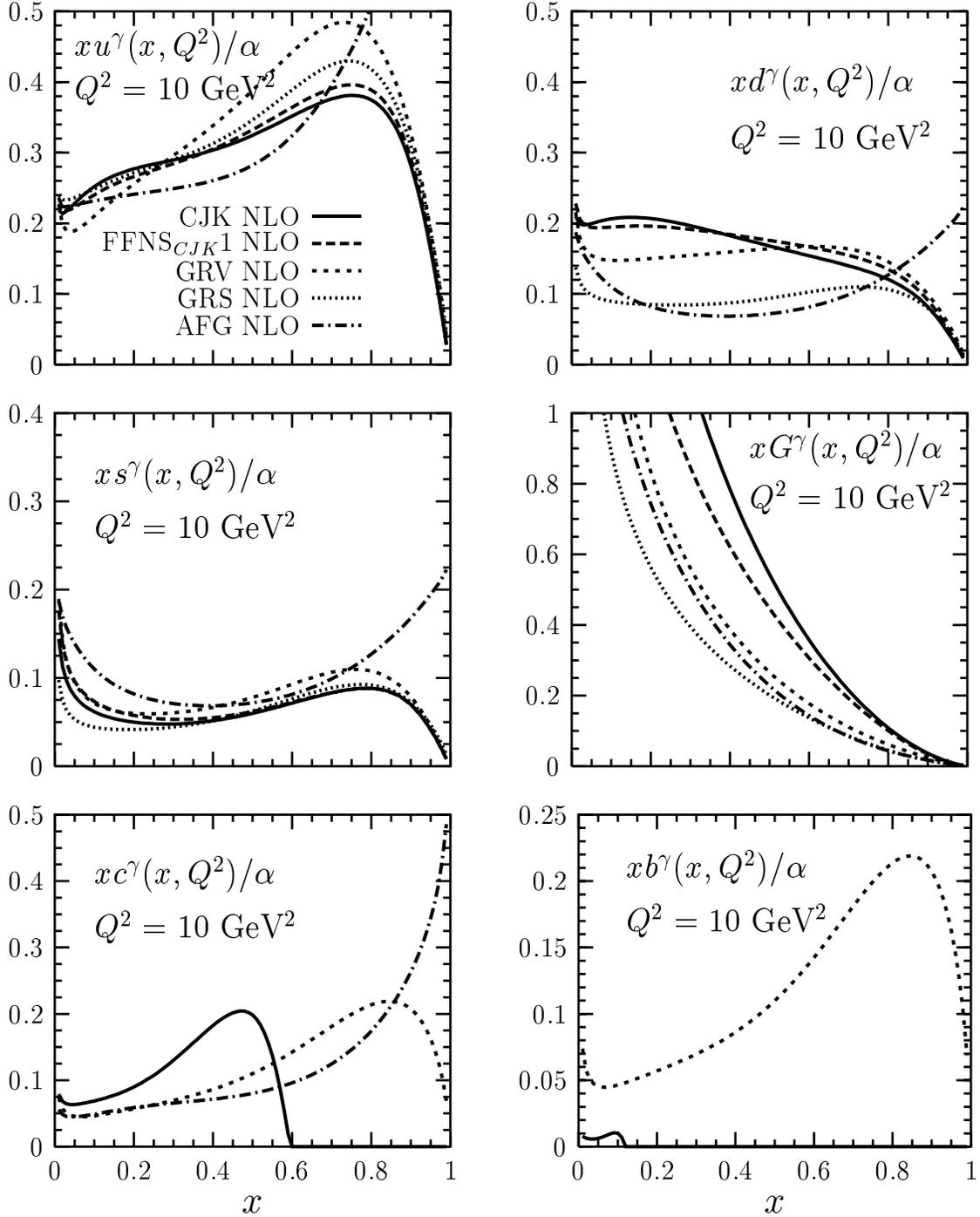}%
\caption{Comparison of the NLO parton densities predicted by CJK NLO and 
FFNS$_{CJK}$1 NLO models and by GRV NLO \cite{grv92}, GRS NLO \cite{grs}
and AFG NLO \cite{afg} parametrizations at $Q^2=10$ GeV$^2$, as a function 
of $x$.}
\label{partonparam}
\end{figure}

\begin{figure}
\begin{center}
\includegraphics[scale=1.0]{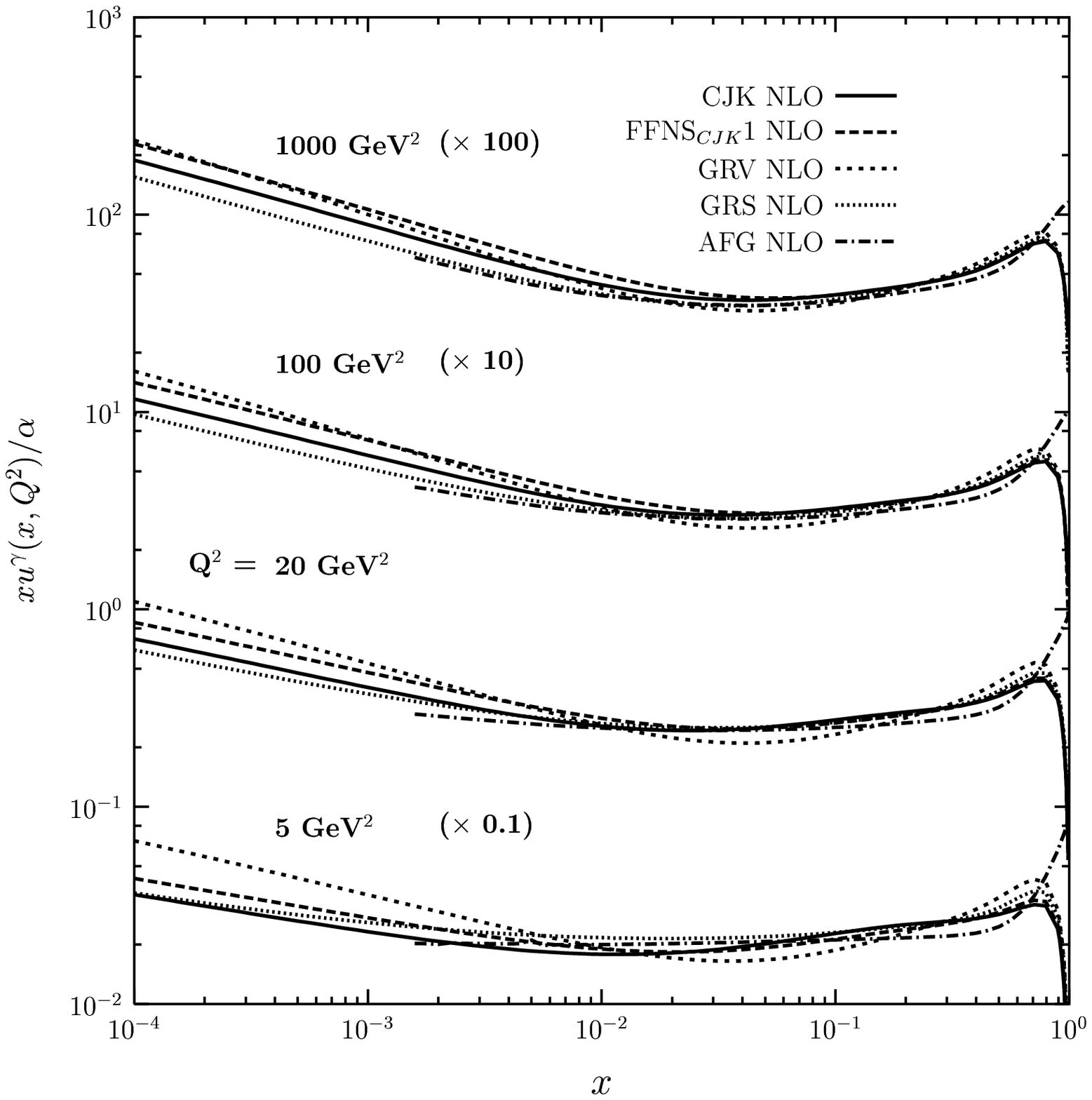}%
\caption{Comparison of the up-quark density at four values of $Q^2$ in the CJK
and FFNS$_{CJK}1$ NLO models with the GRV NLO \cite{grv92}, GRS NLO \cite{grs}
and AFG NLO \cite{afg} densities.}
\label{updens}
\end{center}
\end{figure}

\begin{figure}
\begin{center}
\includegraphics[scale=1.0]{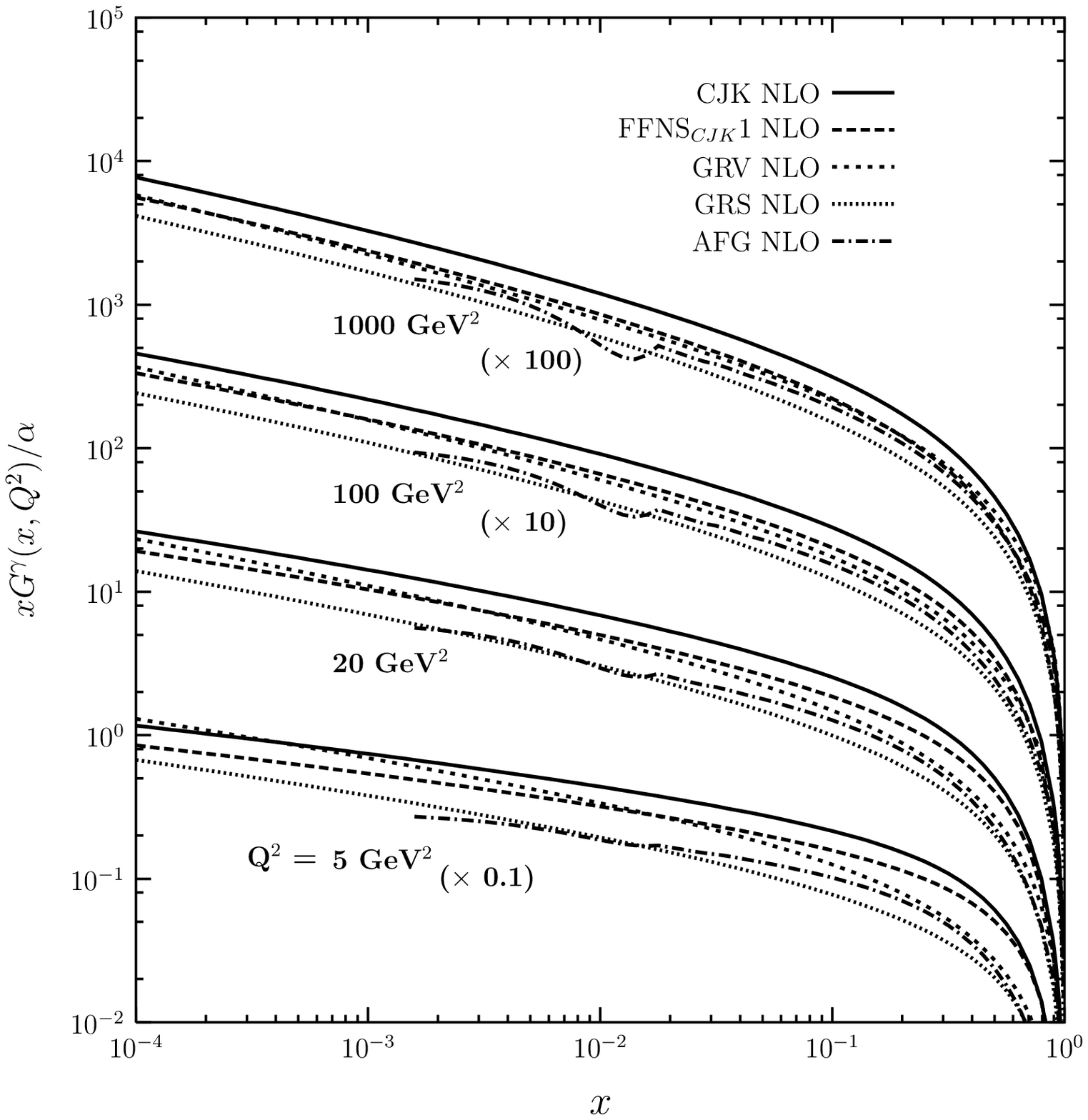}%
\caption{The same as in Fig. \ref{updens} for the gluon density.}
\label{gludens}
\end{center}
\end{figure}

\begin{figure}
\begin{center}
\includegraphics[scale=1.0]{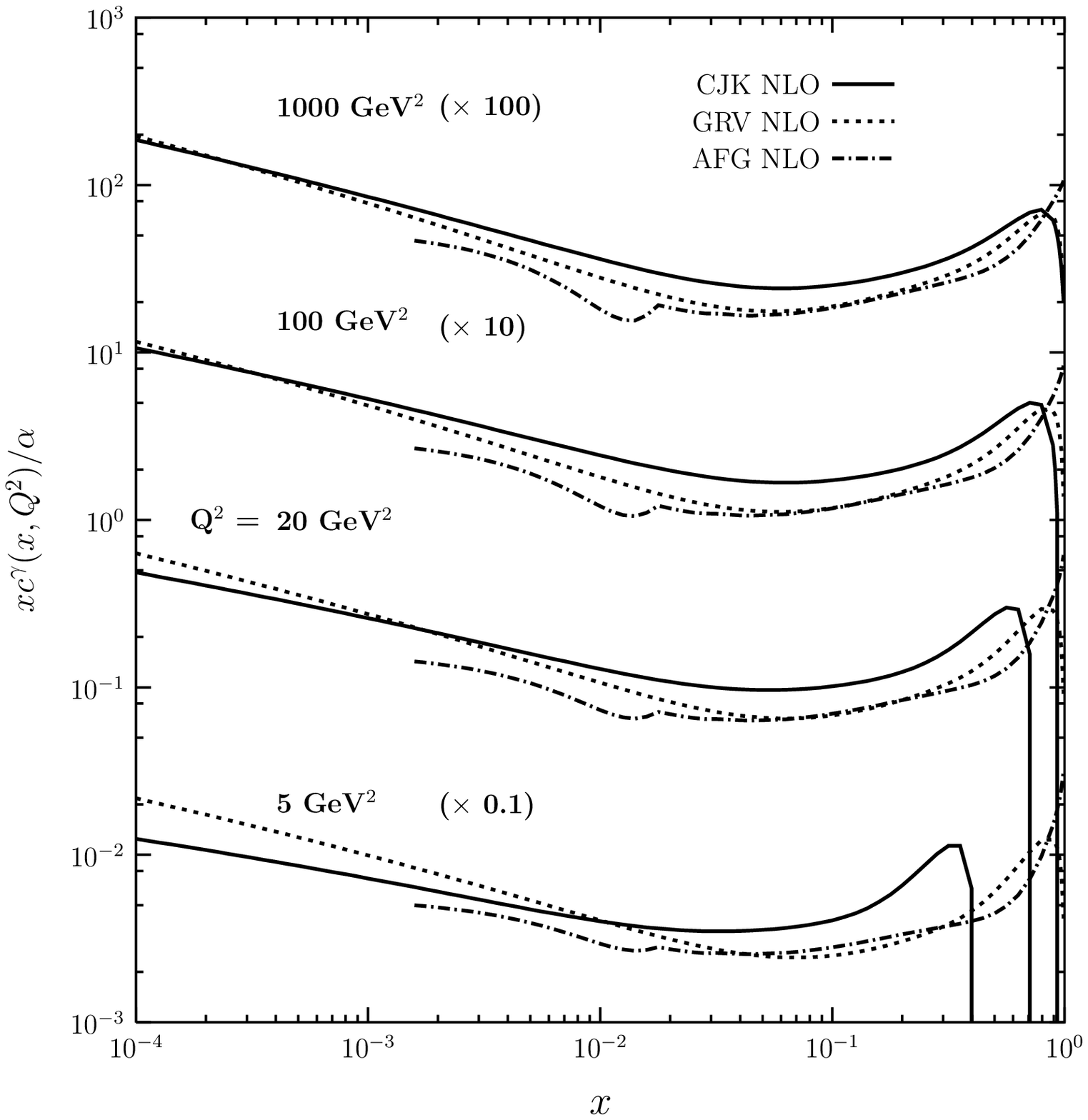}%
\caption{The same as in Fig. \ref{updens} for the charm-quark density.}
\label{chmdens}
\end{center}
\end{figure}

In this section we present the parton densities obtained from the CJK and 
FFNS$_{CJK}1$ NLO fits and compare them with the corresponding distributions 
of the GRV \cite{grv92}, GRS and AFG NLO parametrizations. First, we present 
all parton densities at $Q^2=10$ GeV$^2$ (Fig. \ref{partonparam}). The biggest
difference between our CJK NLO model and others is observed, as expected, for 
the heavy-quark distributions. Unlike for the GRV and AFG NLO parametrizations
(there is no beauty-quark density in the AFG NLO parametrization), our 
$c\gam(x,Q^2)$ and $b\gam(x,Q^2)$ densities vanish not at $x=1$ but, as it 
should be, at the kinematical threshold. Moreover, we notice a huge difference 
between the beauty-quark distributions computed in two models. The GRV NLO 
parametrization predicts the $b\gam(x,Q^2)$ values of the same order as the 
$c\gam(x,Q^2)$. In Fig. \ref{chmdens}, where the charm-quark distributions are 
presented for various $Q^2$ values, apart from different threshold behaviour 
we observe that CJK $c\gam(x,Q^2)$ distribution is the largest of all down to 
very small $x$ value (depending on $Q^2$) where the GRV density becomes larger.
The light-quark and gluon densities calculated in our models and the GRV and 
GRS NLO parametrizations generally have the same shapes but their values 
differ. Main differences are observed in the up-, down-quark and gluon 
distributions. The $u\gam$ densities computed in both our models  are lower 
than in other parametrizations at high $x$. In the Fig. \ref{updens} we see 
differences among predictions of all models at very low $x$. Still we do not 
find a general pattern of those distinctions. In the case of the gluon 
distribution, see Figs. \ref{partonparam} and \ref{gludens}, we observe that 
at all $Q^2$ regions the CJK model gives the largest predictions. At high 
$Q^2$, $G\gam(x,Q^2)$ calculated in the FFNS$_{CJK}1$ NLO model and in GRV NLO 
parametrization become similar. The GRS and AFG NLO predictions lie below all 
other lines in the whole $x$ and $Q^2$ range.

\begin{figure}
\includegraphics[scale=1.0]{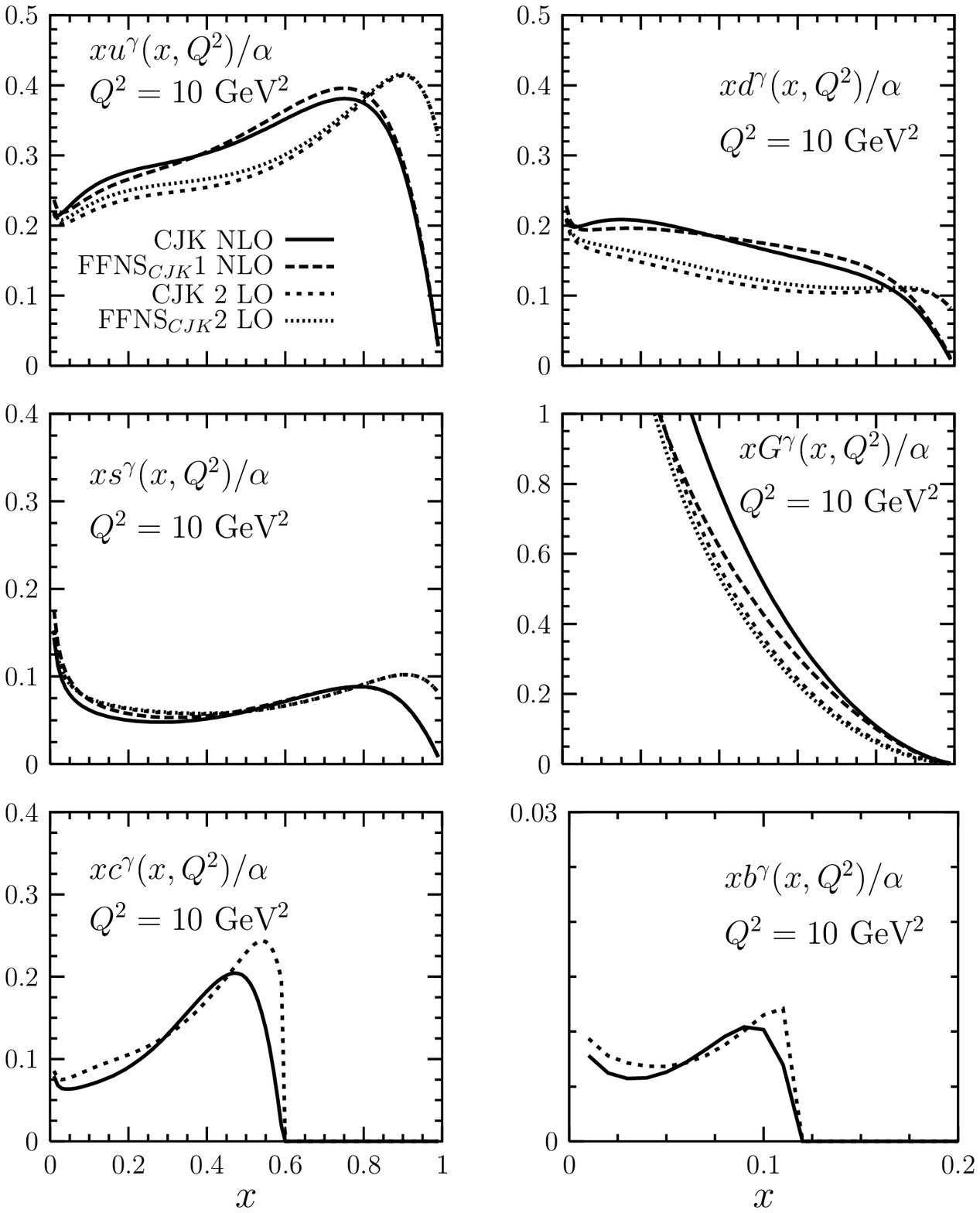}%
\caption{Comparison of the parton densities predicted by the CJK NLO and 
FFNS$_{CJK}$1 NLO models and by corresponding LO models, CJK LO and 
FFNS$_{CJK}$2 LO (\ie including the resolved-photon $\gamma^* G\to h\bar h$ 
contribution) from \cite{cjk} at $Q^2=10$ GeV$^2$, as a function of $x$.}
\label{partonloho}
\end{figure}

Further, we compare our CJK and FFNS$_{CJK}1$ NLO parton distributions with 
the corresponding CJK and FFNS$_{CJK}2$ LO ones. As can be seen in Fig. 
\ref{partonloho}, for $Q^2=10$ GeV$^2$, the NLO models predict higher up- and 
down-quark densities in the medium-$x$ region and steeper increase of the 
gluon distribution. The growth of $G\gam(x,Q^2)$ at low $x$ is especially
fast in the CJK NLO model case while the FFNS$_{CJK}1$ NLO model prediction 
tends to fuse with the LO ones. Moreover, all NLO quark densities vanish more 
rapidly at $x\to 1$ than the LO quark densities.


\subsection{Comparison with $F\gam_{2,c}$}

\begin{figure}
\includegraphics[scale=1.0]{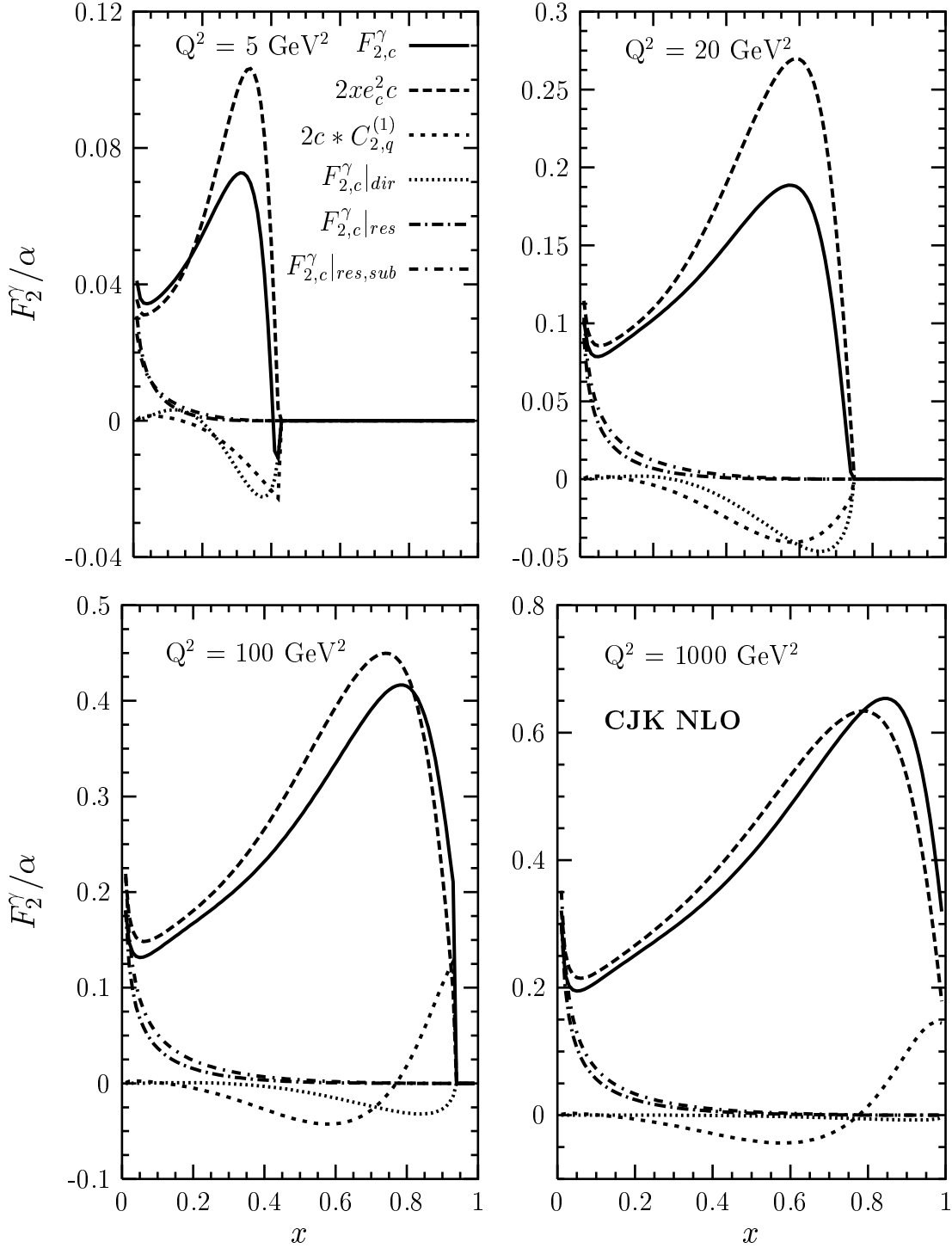}%
\caption{Comparison of various contributions to the photon structure
function $F_{2,c}^{\gamma}(x,Q^2)/\alpha$ in the CJK NLO model for 
$Q^2=5,20,100$ and 1000 GeV$^2$.}
\label{acot}
\end{figure}

In Fig. \ref{acot} we present the CJK NLO predictions for the $F_{2,c}\gam$. 
For $Q^2=5,20,100$ and 1000 GeV$^2$ we compare the individual contributions
included in the model. Analogically to the LO case, see \cite{cjkl} and 
\cite{cjk}, we introduce the following notation
\be
F_{2,c}\gam(x,Q^2) = 2xe_c^2 c\gam(x,Q^2)+ F_{2,c}\gam|_{\mathrm{dir}}(x,Q^2) 
+ F_{2,c}\gam|_{\mathrm{res}}(x,Q^2) - F_{2,c}\gam|_{\mathrm{res,sub}}(x,Q^2) 
+ \left[2c*C_{2,q}^{(1)}\right] (x,Q^2)
\ee
with
\ba
F_{2,c}\gam|_{\mathrm{dir}}(x,Q^2) &=& 3e_c^2\frac{\alpha}{\pi} 
\left[-\beta x(1-x)\frac{4m_c^2}{Q^2} + 
\left(x(1-3x)\frac{4m_c^2}{Q^2}-x^2\frac{8m_c^4}{Q^4} \right)\ln 
\frac{1+\beta}{1-\beta} \right], \nonumber \\
F_{2,c}\gam|_{\mathrm{res}}(x,Q^2) &=& \int_{\chi_c}^1\frac{dy}{y} 
G\gam(y,Q^2)C_{2,G}^{h,(1)}(\frac{x}{y},\frac{Q^2}{m_c^2}), \\ 
F_{2,c}\gam|_{\mathrm{res,sub}}(x,Q^2) &=& \ln\frac{Q^2}{Q_0^2}\cdot 2 
\int_{\chi_c}^1\frac{dy}{y} G\gam(y,Q^2)P_{qG}^{(0)}(\frac{\chi_c}{y}),
\nonumber \\
\left[ 2c*C_{2,q}^{(1)}\right](x,Q^2) &=& 
\int_x^1\frac{dy}{y}(c\gam+\bar c\gam)(y,Q^2)C^{(1)}_{2,q}(\frac{x}{y}).
\nonumber
\ea  
Let us stress that the above direct contribution effectively includes both the 
direct and direct subtraction terms, as discussed in detail in section 
\ref{disacot}. Therefore, it is not the same as the 
$F_{2,c}\gam|_{\mathrm{dir}}(x,Q^2)$ in the LO analyses. 
That way in the NLO analysis we removed the troublesome subtraction term 
$\sim k_q^{(0)}$. Unfortunately, also the $2c*C_{2,q}^{(1)}$ contribution does 
not vanish at the threshold and therefore an additional constraint 
$2c*C_{2,q}^{(1)}=0$ for $\chi_c>1$ is necessary. This term is important in 
the high and medium $x$, its values in the latter range are negative. As 
can be seen in the plot all other terms vanish in the $W\to 2m_c$ threshold. 
The direct term influences the $F_{2,c}\gam$ only in the midium and high $x$ 
at low-$Q^2$ range, where it is negaive. The charm-quark density contribution, 
\ie the term $2xe_c^2 c^{\gamma}(x,Q^2)$, dominates the $F_{2,c}\gam$ in the 
whole kinematically available $x$ range. In the low-$x$ region also the 
resolved-photon contributions increase, but they cancel each other.

\begin{figure}[t]
\includegraphics[scale=0.85]{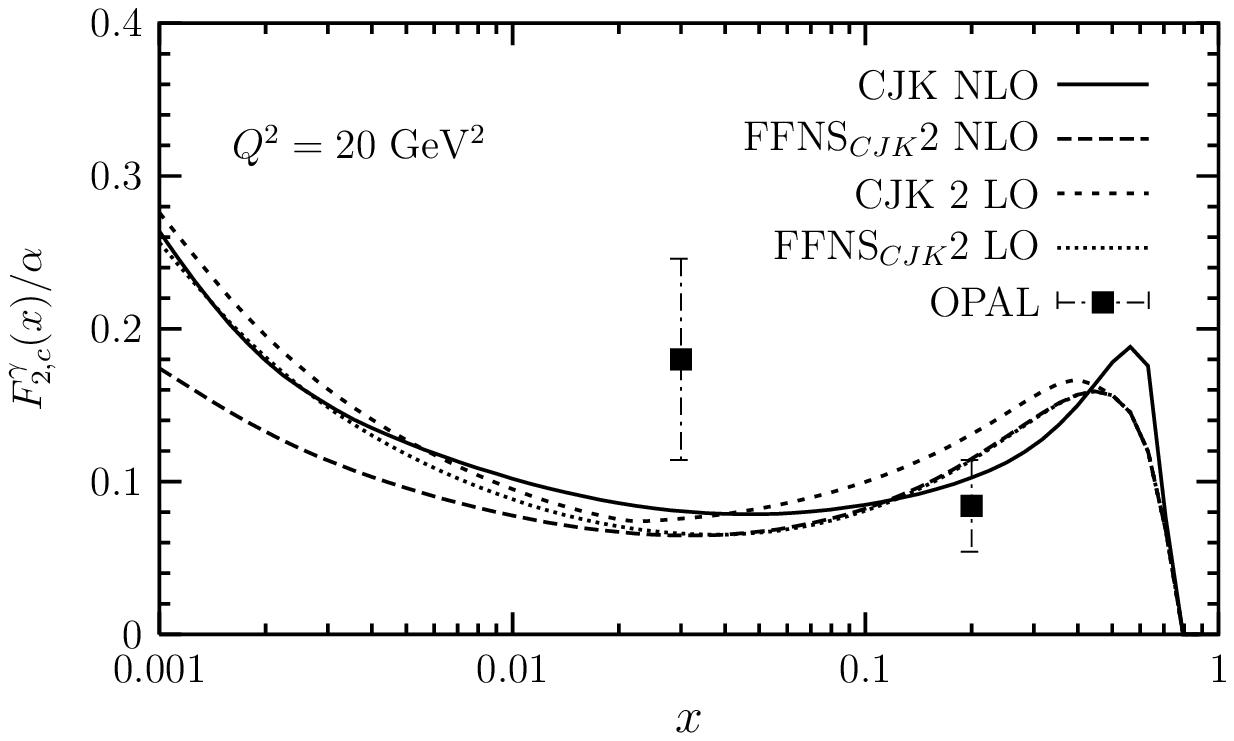}%
\caption{Comparison of the structure function $F_{2,c}^{\gamma}(x,Q^2)/\alpha$
calculated in the CJK and FFNS$_{CJK}$ models for NLO \& LO with the OPAL 
measurement \cite{F2c}.} 
\label{fF2c}
\end{figure}

A good test of the charm-quark contributions is provided by the OPAL 
measurement of the $F_{2,c}\gam$, obtained from the inclusive production of 
$D^{*\pm}$ mesons in the photon--photon collisions \cite{F2c}. The averaged 
$F_{2,c}\gam$ has been determined in the two $x$ bins. These data points are 
compared to the predictions of the CJK and FFNS$_{CJK}$1 NLO models, as well 
as of the corresponding CJK and FFNS$_{CJK}$2 LO models in Fig. \ref{fF2c}. 
The CJK NLO model seems to give the best description of the data, especially 
for the high-$x$ bin. The FFNS type models predict the same $F_{2,c}\gam$ 
at $x>0.03$, while below that value the FFNS$_{CJK}$2 LO predicts higher 
$F_{2,c}\gam$ which finally becomes very similar to the NLO results.


\subsection{Test of importance of $O(\alpha_s^2)$ and $O(\alpha \alpha_s)$ 
terms}

In this section we would like to present the results of the fit of the 
FFNS$_{CJK}$2 NLO model to the data. As discussed in Sections 
\ref{heavyffnssect} and \ref{f2ffnsform}, in this approach we include the 
$O(\alpha_s^2)$ and $O(\alpha \alpha_s)$ contributions to the photon structure 
function:
\ba
&& \frac{1}{x}F_2\gam(x,Q^2)|_{\mathrm{higher\; order}} =
\label{f22} \\
&& \sum_{h(=c,b)}^2 \left\{ \left(\frac{\alpha_s(Q^2)}{2\pi}\right)^2 
\int_{\chi_h}^1\frac{dy}{y} \left[
\sum_{i=1}^{3}(q_i\gam+\bar q_i\gam)(y,Q^2)(e_i^2 C_{2,q}^{(2)}(\frac{x}{y},\frac{Q^2}{m_h^2}) + 
e_h^2 C^{h,(2)}_{2,q}(\frac{x}{y},\frac{Q^2}{m_h^2})) \right. \right. \nonumber \\
&& \left. + e_h^2G\gam(y,Q^2)C_{2,G}^{h,(2)}(\frac{x}{y},\frac{Q^2}{m_h^2}) \right] \left. 
+ \frac{\alpha \alpha_s(Q^2)}{(2\pi)^2} e_h^4 C_{2,\gamma}^{h,(1)}(x,\frac{Q^2}{m_h^2}) 
\right\}. \nonumber
\ea

Alike in the main CJK and FFNS$_{CJK}$1 NLO models our fit was performed 
in two steps. First, we treated $Q_0^2$ as a free parameter in order to obtain 
better agreement with the data, next it was fixed to the estimated value, 
 $Q_0^2 = 0.716$ GeV$^2$, and a final fit was done. Its results are presented
in table \ref{tparamhigh}, all errors were obtained from \textsc{Minos} 
requiring $\Delta \chi^2 = 1$.

\begin{table}[h]
\begin{center}
\renewcommand{\arraystretch}{1.5}
\begin{tabular}{|c|@{} p{0.1cm} @{}|c|c|@{} p{0.1cm} @{}|c|c|c|@{} p{0.1cm} @{}|c|c|}
\hline
 NLO model         && $\chi^2$ & $\chi^2/_{DOF}$ && $\kappa$ & $\alpha$ & $\beta$ && $N_v$ & $\tilde N_G$ \\
\hline
\hline
 FFNS$_{CJK}$2 && 252.2 & 1.31 && 1.892$^{+0.101}_{-0.090}$ & 0.527$^{+0.089}_{-0.080}$ & 0.966$^{+0.387}_{-0.333}$ && 0.797 & 2.727 \\
\hline
\end{tabular}
\caption{\small The $\chi^2$ and parameters of the fit for the FFNS$_{CJK}$2 
NLO model with assumed $Q_0^2 = 0.716$ GeV$^2$. The $\alpha$, $\beta$ and 
$\kappa$ symmetric parabolic errors obtained from \textsc{Minos} requiring 
$\Delta \chi^2 = 1$.}
\label{tparamhigh}
\end{center}
\end{table}

First let us compare the results of the two FFNS$_{CJK}$ NLO models, given in
tables \ref{tparam} and \ref{tparamhigh}, respectively. We notice that higher 
$\chi^2/_{DOF}$ is obtained in the model including additional terms. Next, the
input scale and the parameters obtained in both approaches are very similar. 
Largest distinction is observed between the $\kappa$ values and can be 
explained by the small difference between the $Q_0^2$ values (0.716 and 0.786 
GeV$^2$, respectively). Namely, when we start the evolution of the parton 
densities at lower input scale then smaller input distributions are required 
in order to obtain the same results at the given $Q^2$.

Further, we compare results computed using four various FFNS$_{CJK}$ models, 
the NLO ones described in this work as well as the FFNS$_{CJK}1$ \& 2 LO 
models, presented in \cite{cjk}. The main differences among the four models, 
apart from the different order of the QCD DGLAP equations applied, lie in the 
heavy-quark contributions to $F_2\gam$. Those differences are summerized in 
table \ref{ffnsdiff} where all included or discluded terms proportional to 
Wilson-coefficient functions are listed.

\begin{table}[h]
\begin{center}
\renewcommand{\arraystretch}{1.5}
\begin{tabular}{|c|@{} p{0.1cm} @{}|c|c|c|c|c|c|}
\hline
 FFNS$_{CJK}$ model && $C_{2,\gamma}^{h,(0)}$ & $C_{2,G}^{h,(1)}$ & $C_{2,\gamma}^{h,(1)}$ & $C_{2,G}^{h,(2)}$ & $C_{2,q}^{h,(2)}$ & $C_{2,q}^{(2)}$ \\
\hline
\hline
 LO 1  && + & - & - & - & - & - \\
\hline
 LO 2  && + & + & - & - & - & - \\
\hline
 NLO 1 && + & + & - & - & - & - \\
\hline
 NLO 2 && + & + & + & + & + & + \\
\hline
\end{tabular}
\caption{\small The heavy-quark terms proportional to the Wilson coefficient 
functions contributing (or not) to the heavy-quark structure function in 
various FFNS$_{CJK}$ models.}
\label{ffnsdiff}
\end{center}
\end{table}

\begin{figure}[b]
\includegraphics[scale=0.85]{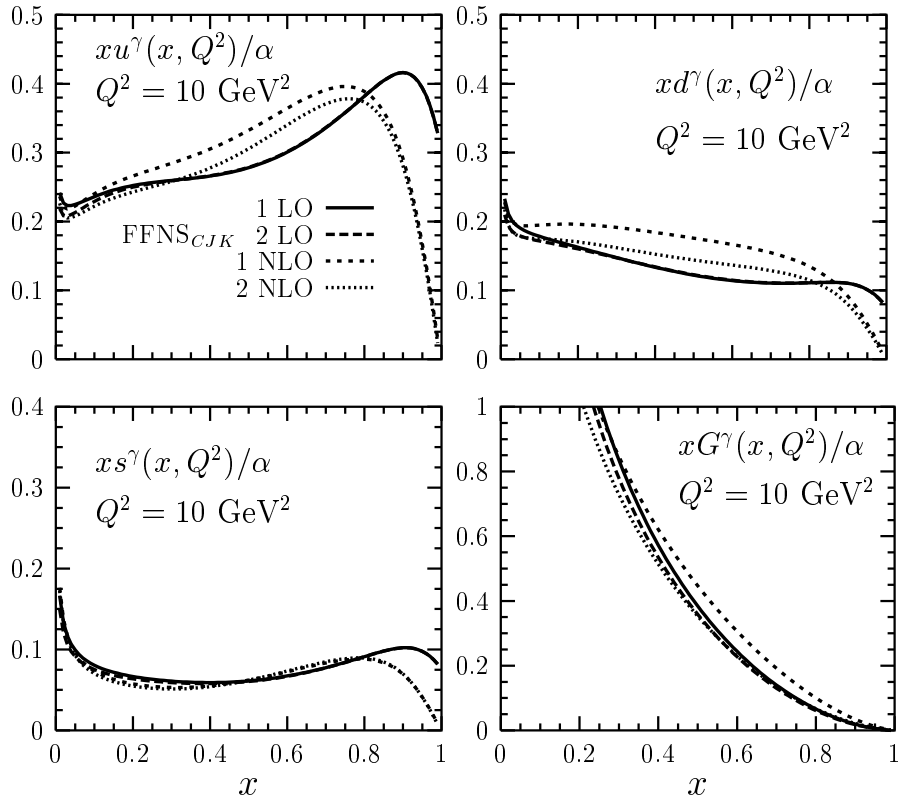}%
\caption{Comparison of the parton densities predicted by the FFNS$_{CJK}$1 \& 2
NLO models and by the corresponding FFNS$_{CJK}$1 \& 2 LO models presented in 
\cite{cjk} at $Q^2=10$ GeV$^2$, as a function of $x$.}
\label{partonffns}
\end{figure}

\begin{figure}
\hskip -0.5cm
\includegraphics[scale=1.0]{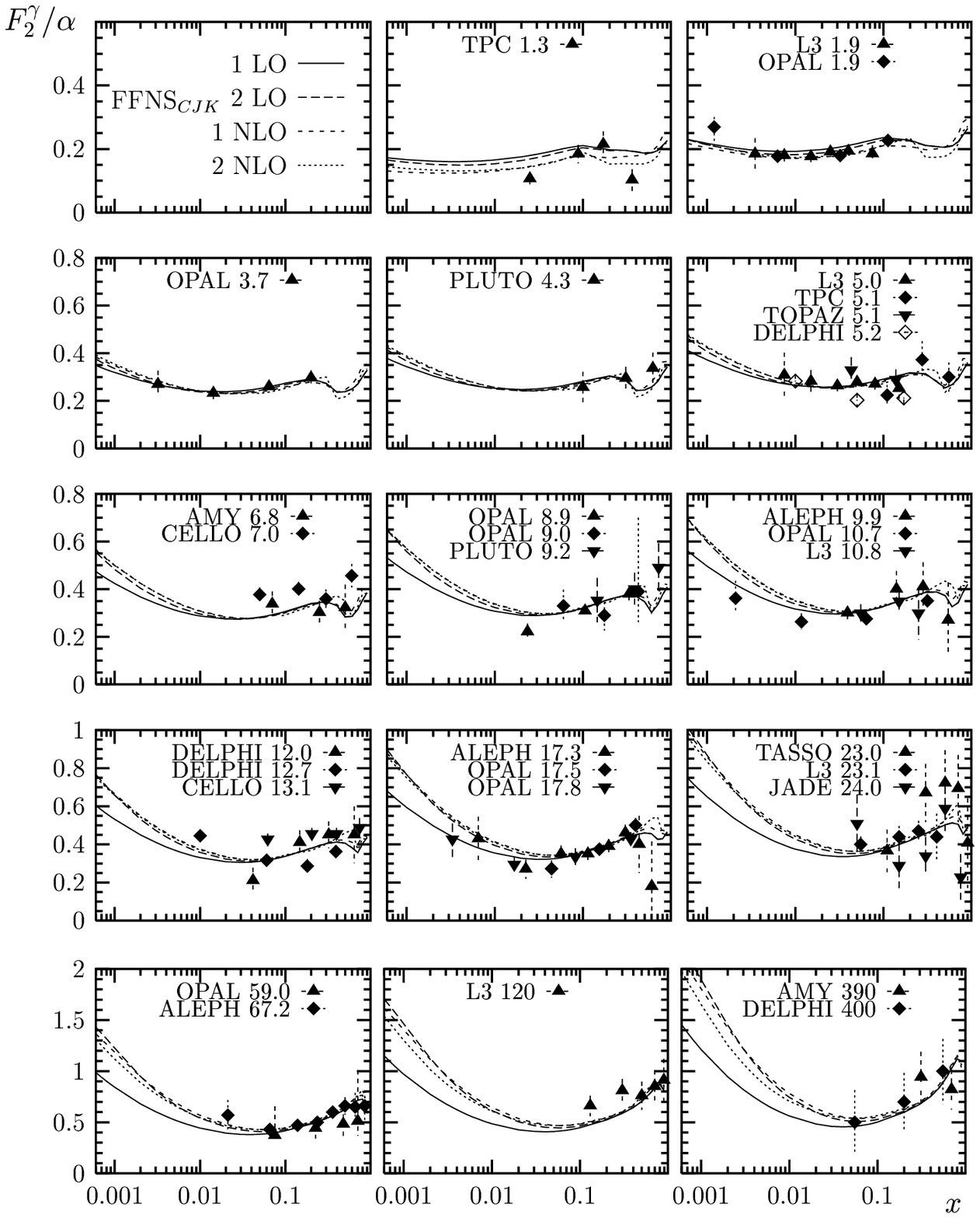}%
\vskip -0.2cm
\caption{Predictions for the $F_2^{\gamma}(x,Q^2)/\alpha$ for the FFNS$_{CJK}$
models, as in Fig. \ref{partonffns}, compared with the corresponding LO fits 
from \cite{cjk} and with the experimental data \cite{CELLO}--\cite{HQ2}, for 
small and medium $Q^2$ as a function of $x$ (logarithmic scale). If a few 
values of $Q^2$ are displayed in the panel, the average of the smallest and 
biggest $Q^2$ was taken in the computation.}
\label{fitffns1}
\end{figure}

\begin{figure}
\hskip -0.5cm
\includegraphics[scale=1.0]{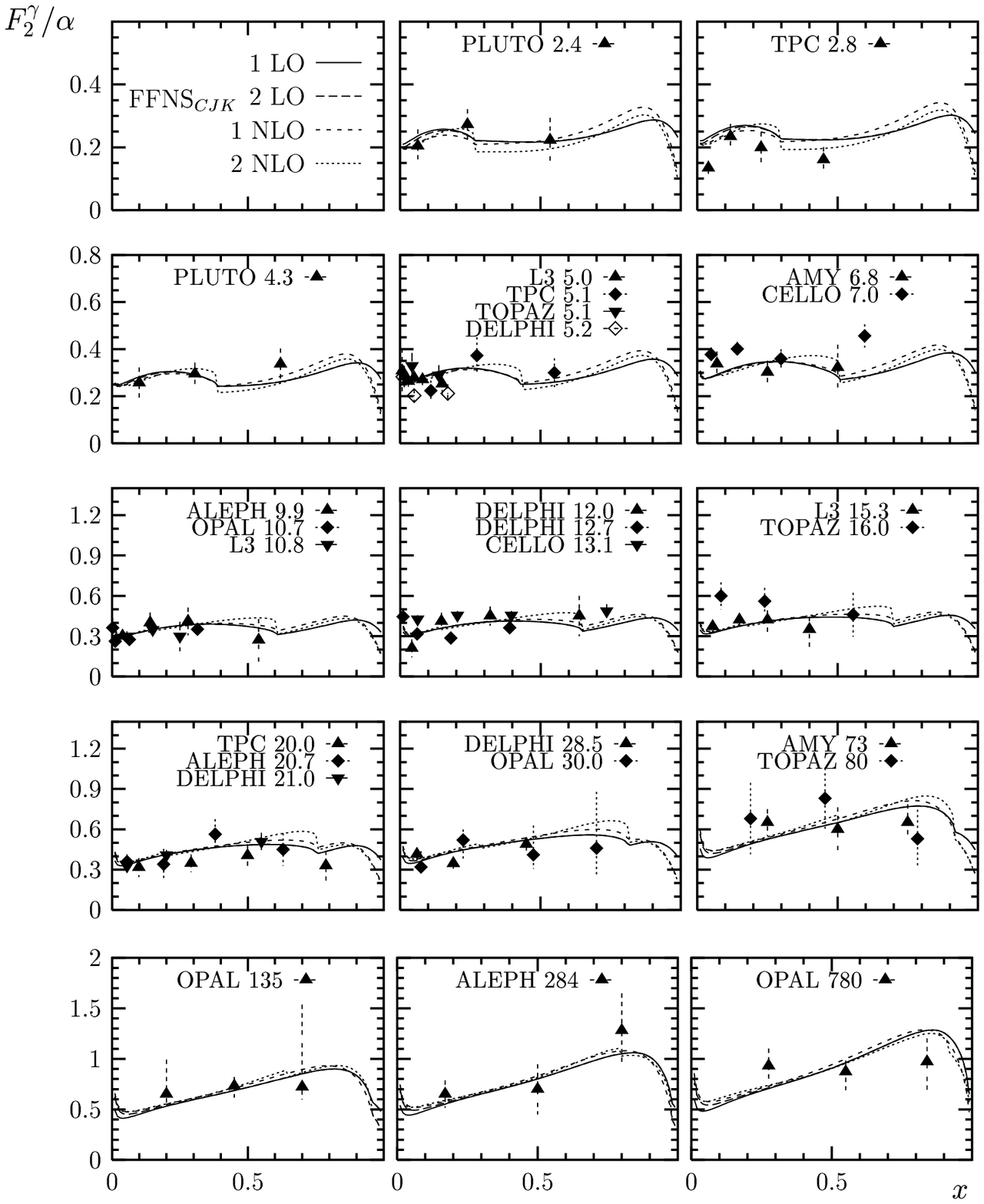}%
\caption{The same as in Fig. \ref{fitffns1} for a linear scale in $x$.}
\label{fitffns2}
\end{figure}

In Fig. \ref{partonffns} we compare the parton densities computed in the
FFNS$_{CJK}$ models obtained in LO and NLO. We observe that two LO models 
predict nearly the same parton distributions which begin to differ in the very 
low-$x$ region (the larger differences at $x$ below 0.1 are not shown). The 
distinctions between the pairs of LO and NLO lines are large at high $x$ where,
as discussed before, the NLO distributions decrease much earlier. The maximal 
values of the NLO $u\gam$ and $s\gam$ distributions are reached at smaller $x$,
$x \sim 0.75$ than in the LO case where the corresponding $x$ value is 
$x\sim 0.95$. Moreover, there exist important differences between the two NLO 
fits, observed especially in the up- and down-quark distributions, which 
indicate the importance of the higher order $O(\alpha_s^2)$ and 
$O(\alpha \alpha_s)$ contributions to \fund at medium $x$. Smaller difference 
between the LO results seen in Fig. \ref{partonffns} may be explained by the 
fact that in that case the models differ only by the resolved-type 
$\sim C_{2,G}^{h,(1)}$ contribution, which plays important role only in the 
low-$x$ region. Very similar conclusion may be drawn from the comparison of 
the \fun distributions plotted in Figs. \ref{fitffns1} and \ref{fitffns2}. 
Alike in the parton densities case the LO \fun distributions differ only in 
the very low-$x$ region, see Fig. \ref{fitffns2}, while the NLO ones mostly at
high $x$.


\subsection{Influence of the DELPHI LEP2 data}

\begin{figure}
\includegraphics[scale=1.0]{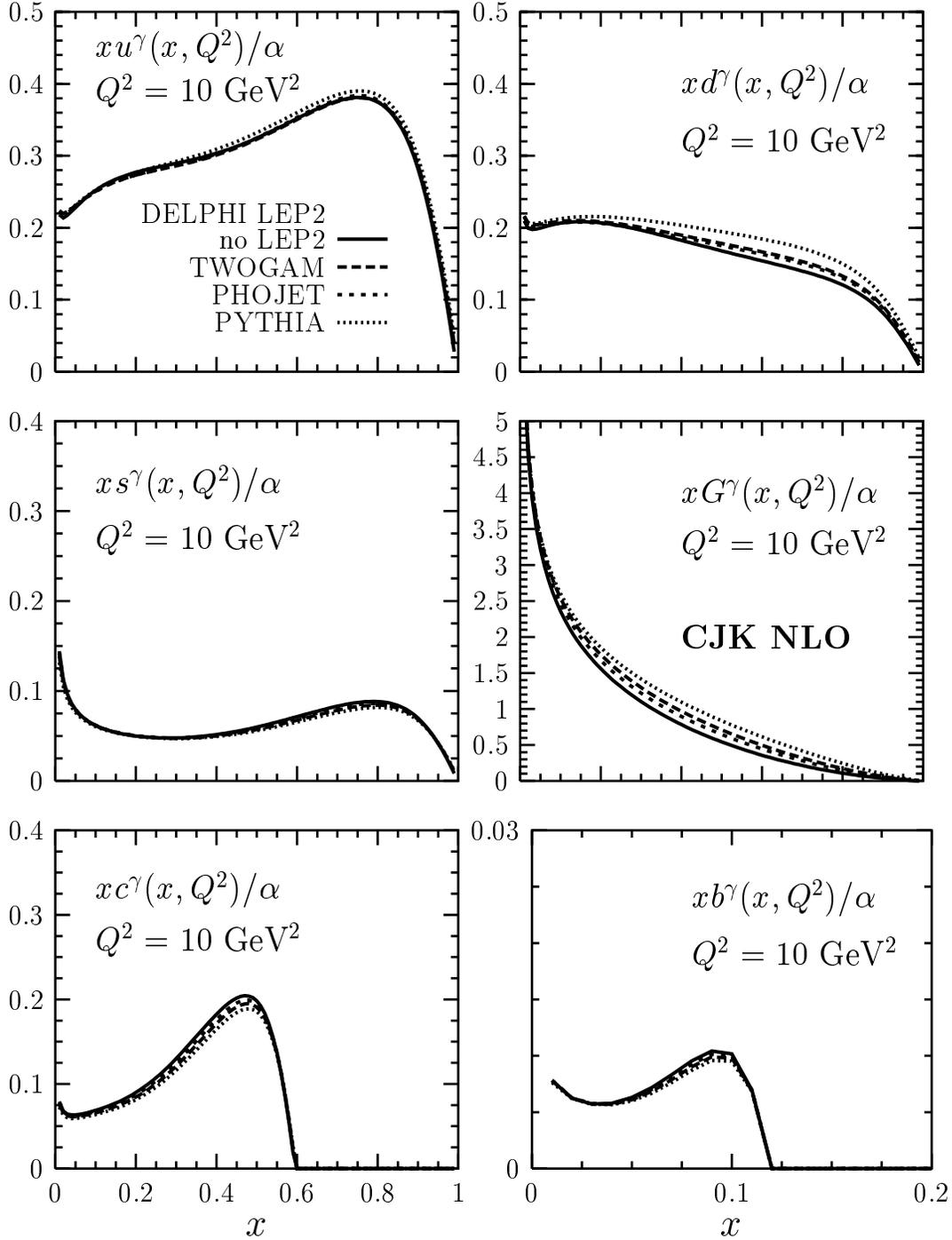}%
\caption{Comparison of the parton densities predicted by the test fits obtained
with the CJK NLO model using various sets of the DELPHI LEP2 data 
\cite{DELPHInew} at $Q^2=10$ GeV$^2$, as a function of $x$.}
\label{partondelphi}
\end{figure}

\begin{figure}
\hskip -1cm
\includegraphics[scale=1.0]{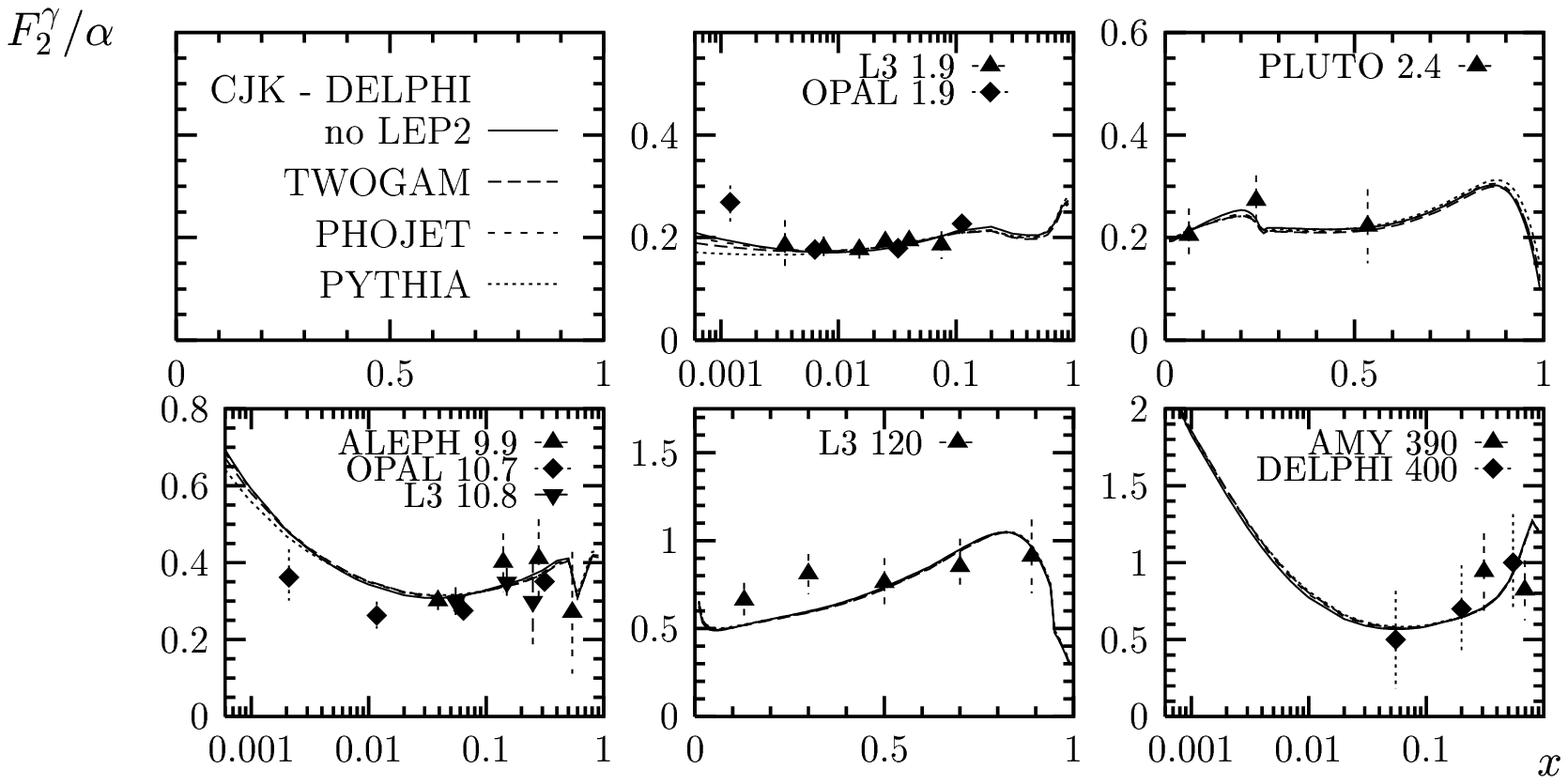}%
\caption{Predictions for the $F_2^{\gamma}(x,Q^2)/\alpha$ for the CJK NLO model
 obtained in the test fits using various sets of the DELPHI LEP2 data 
\cite{DELPHInew}.}
\label{fitdelphi}
\end{figure}

Finally we tested the possible influence of the recent DELPHI data 
\cite{DELPHInew}, which were not included in the final fits, on the results of 
our models. We performed test fits for the CJK NLO model including separately 
each of the sets of the LEP2 data obtained with TWOGAM \cite{TWOGAM}, PHOJET 
\cite{PHOJET} and PYTHIA \cite{PYTHIA} Monte Carlo generators and presented in 
\cite{DELPHInew}. As can be seen in Figs. \ref{partondelphi} and 
\ref{fitdelphi} for the choice of the \fun distributions and the parton 
densities computed at $Q^2=10$ GeV$^2$, respectively, the results of that fits 
are very similar to the expectations of the final CJK NLO fit (denoted as 
``no LEP2''). Still, the $\chi^2/_{DOF}$ obtained in the test fits are much 
worse, see table \ref{tparamdelphi}.

\begin{table}[htb]
\begin{center}
\renewcommand{\arraystretch}{1.5}
\begin{tabular}{|c|@{} p{0.1cm} @{}|c|c|c|@{} p{0.1cm} @{}|c|@{} p{0.1cm} @{}|c|c|c|@{} p{0.1cm} @{}|c|c|}
\hline
set && $\chi^2$ & \# of points & $\chi^2/_{DOF}$ \\
\hline
\hline
TWOGAM && 307.4 & 209 & 1.50 \\
\hline
PHOJET && 312.0 & 207 & 1.54 \\
\hline
PYTHIA && 341.1 & 209 & 1.66 \\
\hline
\end{tabular}
\caption{\small Number of experimental points used in the test fits for the 
CJK NLO model and the resulting $\chi^2$.}
\label{tparamdelphi}
\end{center}
\end{table}



\section{Disscusion and conclusions}

A new analysis of the radiatively generated parton distributions in the real
photon based on the NLO DGLAP equations is presented. An updated set of the
\fun data including recent ALEPH as well as new preliminary DELPHI measurements
has been used to perform three global fits. Our models base on two schemes, 
the Fixed Flavour-Number Scheme and Variable Flavour-Number Scheme 
(ACOT$(\chi)$) which were already applied in our former LO analysis, see
\cite{cjkl}-\cite{jank}. We observe that in comparison to the LO models 
higher input scale of the NLO DGLAP evolution is required in order to 
satisfactory describe the data. Further, unlike in the LO case we obtained 
lower $\chi^2/_{DOF}$ in the FFNS type approach than in the CJK model. 
The test fits showed that a change of the simple form of the input parton
distributions applied in \cite{cjkl,cjk,jank} and in this work does not 
lead to the improvement of the fits.

Predictions of the FFNS$_{CJK}1$ and CJK NLO fits are compared with the \fun 
data, other parametrizations and corresponding LO models. Both models describe
very well the $Q^2$ evolution of the \fun, averaged over various $x$-regions.
The $F_{2,c}\gam$ calculated in the CJK NLO model gives best agreement with 
the data. 

Next, we check the difference between the results obtained in the FFNS$_{CJK}1$
NLO model and the FFNS$_{CJK}2$ NLO model which includes additional 
$O(\alpha_s^2)$ and $O(\alpha \alpha_s)$ contributions to \fund. Those higher 
order terms prove to be of importance. Therefore, further studies of the CJK 
model including terms of the same order are required.

Finally, we examine the influence on the results of our models of the various 
sets of the LEP2 data presented in the recent preliminary results of the 
DELPHI Collaboration but discluded from final fits. Test fits show very small 
change of the \fun and parton distributions obtained using these data 
comparing to the final fits but large deterioration of their quality.

Fortran parametrization programs for our CJK and FFNS$_{CJK}$ NLO
models can be obtained from the web-page \cite{webprog}.


\section*{Acknowledgements}

P.J. would like to thank E.Laenen for making accesible the formulae for
the Wilson coefficients and for his comments. Moreover, P.J. is grateful to
I.Schienbein for his help in obtaining correct NLO DGLAP evolution programs.
We want to thank R.Gokieli and K.Doroba for their comments on the recent DELPHI
data. Finally P.J. is grateful to M.Aurenche and M.Fontannaz for the AFG 
parametrization program, R.Nisius for all his remarks concerning the 
experimental data and A.Zembrzuski for his comments.

This work was partly supported by the European Community's Human Potential 
Programme under contract HPRN-CT-2000-00149 Physics at Collider and 
HPRN-CT-2002-00311 EURIDICE. FC also acknowledges partial financial
support from Ministerio de Ciencia y Tecnolog{\'\i}a under project 
FPA2003-09298-c02-01 and Junta de Andaluc{\'\i}a under project FQM 330. This 
work was partially supported by the Polish Committee for Scientific Research, 
grant~no.~1~P03B~040~26 and project 
no.~115/E-343/SPB/DESY/P-03/DWM517/2003-2005.


\section*{Appendix: The Wilson-coefficient functions}

Lists of all Wilson-coefficient functions used in this analysis are given
in tables \ref{WCL} and \ref{WCH}. The Wilson coefficients $C_{2,q}^{(1)}$, 
$C_{2,G}^{(1)}$ and $C_{2,\gamma}^{(0)}$ have been first calculated in the 
operator product expansion \cite{OPE} in the Mellin space in the articles 
\cite{C2qg} and \cite{C2gam}, respectively. In the $(x,Q^2)$ space the 
$C_{2,\gamma}^{(1)}$ in the \ms factorization scheme reads (see for instance 
\cite{grs})
\be
C_{2,\gamma}^{(1)}(x) = 6 \left[-1+8x(1-x)+\left(x^2+(1-x)^2\right)\ln \frac{1-x}{x}\right].
\ee

The lowest order heavy-quark coefficient $C_{2,\gamma}^{h,(0)}$, given by the 
Bethe-Heitler $\gamma^* + \gamma \to q+\bar q$ process, has been first 
presented in \cite{BetheHeitler}. It has the following form:
\ba
C_{2,\gamma}^{h,(0)}(x,\frac{Q^2}{m_h^2}) &=& 6\frac{\alpha}{\pi}e_h^4
\beta \bigg[ -1+8x(1-x)-x(1-x)\frac{4m_h^2}{Q^2} \bigg] \\
&& +\ln \left(\frac{1+\beta}{1-\beta}\right)\bigg[x^2+(1-x)^2
+x(1-3x)\frac{4m_h^2}{Q^2}-x^2\frac{8m_h^4}{Q^4}\bigg], \nonumber
\label{bhform}
\ea
with $\beta = \sqrt{1-\frac{4m_h^2x}{(1-x)Q^2}}$.

When we neglect the heavy-quark mass in some parts of the coefficient
$C_{2,\gamma}^{h,(0)}$ we can rewrite it as in Eq. (\ref{c2qmsm})
\ba
C_{2,\gamma}^{h,(0),\msm}(x,\frac{Q^2}{m_h^2}) 
&\approx& C_{2,\gamma}^{(0),\msm}(x) 
+ e_h^4 2k_q^{(0),\msm}(x)\ln \frac{Q^2}{m_h^2} \\
&+& 6e_h^4 \left[-\beta x(1-x)\frac{4m_h^2}{Q^2} +
\left(x(1-3x)\frac{4m_h^2}{Q^2}-x^2\frac{8m_h^4}{Q^4} \right)\ln 
\frac{1+\beta}{1-\beta} \right], \nonumber
\ea
where the direct photon-quark splitting function is
\be
k_q^{(0)} = 3 [x^2+(1-x)^2].
\ee

The $C_{2,G}^{h,(1)}$ function related to the $\gamma^* + G \to q+\bar q$ 
process reads
\ba
C_{2,G}^{h,(1)}(x,\frac{Q^2}{m_h^2}) &=& 
\frac{1}{6}C_{2,\gamma}^{h,(0)}(x,\frac{Q^2}{m_h^2}) \\
&=& \frac{\alpha}{\pi}e_h^4 \beta 
\bigg[ -1+8x(1-x)-x(1-x)\frac{4m_h^2}{Q^2} \bigg] \nonumber \\
&& +\ln \left(\frac{1+\beta}{1-\beta}\right)\bigg[x^2+(1-x)^2
+x(1-3x)\frac{4m_h^2}{Q^2}-x^2\frac{8m_h^4}{Q^4}\bigg]. \nonumber
\ea

Finally, the higher order heavy-quark coefficients, $C_{2,\gamma}^{h,{(1)}}$, 
$C_{2,G}^{(2)}$, $C_{2,q}^{h,{(2)}}$ and $C_{2,q}^{(2)}$ are calculated in
the serious of publications \cite{LRSN}. They are given in the form of a two
dimensional array. Those $O(\alpha_s^2)$ and $O(\alpha \alpha_s)$ order Wilson 
coefficients are given as
\be
C_{2,i}^{h,{(k)}}(x,m_h^2,Q^2,\mu^2) = C_{2,i}^{h,{(k)}}(x,\frac{Q^2}{m_h^2}) 
+ \bar C_{2,i}^{h,{(k)}}(x,\frac{Q^2}{m_h^2})\ln \frac{\mu^2}{m_h^2},
\ee
where $\mu^2$ is the mass factorization scale. In our calculations we simplify
the above formula by choosing $\mu^2 = m_h^2$.



\clearpage

\clearpage

\clearpage

\clearpage

\clearpage

\clearpage

\clearpage

\clearpage

\clearpage

\clearpage

\clearpage

\clearpage

\clearpage

\clearpage

\clearpage

\clearpage

\clearpage

\clearpage

\clearpage

\end{document}